# Transformation-mediated twinning governs plasticity in body-centered cubic nanocrystals under extreme loading

Jan Očenášek[1], Jesper Byggmästar[2], Guanying Wei [2],
Javier Domínguez[3], and Jorge Alcalá[4*]

[1] New Technologies Research Centre, University of West Bohemia in Pilsen,
30614 Plzeň, Czech Republic
[2]Department of Physics, University of Helsinki, Helsinki, FIN-00014, Finland.
[3] National Centre for Nuclear Research, NOMATEN CoE, Andrzeja Sołtana 7,
05-400 Otwock-Świerk, Poland
[4]Department of Materials Science and Metallurgical Engineering, ETSEIB.
Universitat Politècnica de Catalunya, 08028 Barcelona, Spain.

## Abstract

Plasticity in body-centered cubic (BCC) nanocrystals is often associated with twin nucleation phenomena under extreme loading conditions. Here, we reveal unconventional twinning pathways that operate at the intersection of crystal plasticity and structural phase transitions. We show that the classical shear-driven twinning mode becomes progressively suppressed with increasing pressure, giving rise to transformation-mediated twinning pathways involving transient HCP or FCC phases. In BCC Fe, Ta, and Nb nanocrystals of moderate elastic stiffness, plasticity is consistently initiated by an elastic instability that triggers a dual-shuffle process mediated by stable or metastable hexagonal closed-packed (HCP) phases. This pathway operates independently of the characteristic $\{112\}$ twin boundary planes and is driven by compression, challenging the conceptual paradigm for metal plasticity in which plastic deformation arises from shear stresses resolved on specific planes. By contrast, in the archetypal elastically stiffer BCC Mo and W nanocrystals, plastic deformation proceeds via two alternative twinning pathways associated with shear-driven elastic instabilities mediated by highly-distorted face-centered cubic (FCC) phases. Comprehensive analyses of the energy landscapes to the competing nanoscale twinning modes provide mechanistic insight into their activation, establishing a unified framework for transformation-mediated twinning in BCC nanocrystals across a broad range of loading conditions.

*Submitted to Communications Materials*

* Corresponding author: jorge.alcala@upc.edu (Jorge Alcalá)

## 1. Introduction

The structural reliability of strong body-centered cubic (BCC) metallic nanocrystals and nanoscale devices is governed by the spontaneous nucleation of twins and dislocations at free surfaces, which triggers plastic collapse ([1–12]). These defect processes are largely dependent on crystal structure, crystallographic orientation, and loading configuration, and typically arise at stresses far exceeding those required to mobilize preexisting dislocations in bulk crystals.

A distinctive feature to the initiation of plasticity in BCC nanocrystals concerns the asymmetry of the flow stresses under tension and compression, commonly attributed to the competition between shear-driven nucleation processes of twins and dislocations [13–16]. Although dislocation nucleation is expected to prevail when the Schmid factor (*SF*) acting on the critical twinning systems is minimized, a systematic understanding of how extreme loading conditions fundamentally shape the twinning pathways across different BCC metals is lacking. Along these lines, recent studies have reported the emergence of an unconventional anti-twinning process at high stresses [17,18]. In contrast to classic twinning, which proceeds through the glide of partial (twinning) dislocations along $\langle 11\bar{1}\rangle$ directions of $\{112\}$ planes, anti-twinning deformations remain controversial, as they rely on partial dislocation motion along high-energy $\langle\bar{1}\bar{1}1\rangle$ directions under substantially elevated stresses. To mitigate the need for such energetically unfavorable pathways, alternated twinning/anti-twinning processes have also been proposed [18,19].

Further challenges to the understanding of nanoscale plasticity in strong BCC metals arise from the interplay between deformation twinning and stress-induced phase transitions. This interplay reflects the coupling between deformation twinning and the onset of displacive transformations from the parent BCC lattice into either face-centered cubic (FCC) or hexagonal (HCP) structures at high stresses, as reported through Transmission Electron Microcopy (TEM) observations and Molecular Dynamics (MD) simulations of shocked nanocrystalline metals ([20–26]).

Here, we present a mechanistic framework that elucidates the onset of plasticity in BCC nanocrystals and extends the current knowledge based on classic twin nucleation processes. Using advanced MD and DFT simulations with quantum-level accurate atomic potentials, we reveal the atomic-scale mechanisms leading to the formation of transient HCP and FCC phases that mediate twin nucleation and growth in archetypal Ta, Fe, Nb, W, and Mo nanocrystals. These pathways either supersede or synergistically interact with the classical twinning mode in BCC metals across a broad range of loading conditions, including tension, compression, indentation, and crack-tip-induced plasticity. Whereas classical twinning operates at smaller stresses in crystals containing preexisting defects, the transformation-mediated twinning pathways revealed here activate at the much higher stresses associated with the initiation of plasticity in minuscule length scales.

## 2. Results and discussion

### 2.1. The classic twinning pathway in tensile loaded nanocrystals

Figure 1 illustrates the classic twinning mode that governs the initiation of plasticity in our simulations of nanocrystals subjected to tensile loads along $\langle 100\rangle$ directions. This classical twinning mode is fully consistent with that attained at greater material scales under smaller stresses, involving the sequential glide of twinning dislocations with Burgers vector $b = a/6\,\langle 11\bar{1}\rangle$ —where $a$ is the lattice parameter— on habit $\{112\}$ $K_1$-planes along the lower energy easy-glide $\eta_1 = \langle 11\bar{1}\rangle$ directions (see Fig. 1(a) and Refs. [27,28]). A single dislocation passage then causes twin propagation by the magnitude $|b|$, while simultaneously thickening the twin by one atomic spacing normal to the glide $\{112\}$ plane. In $\langle 100\rangle$-oriented nanocrystals under tension, twin growth is driven by the resolved shear stress $\tau_r$ on the twinning $\{112\}\langle 11\bar{1}\rangle$ systems for which the Schmid factor, $SF_t$, $\approx$ 0.47 substantially exceeds that on the anti-twinning $\{112\}\langle\bar{1}\bar{1}1\rangle$ systems, $SF_{\mathrm{a}}$ ≈ 0.24. Similar trends are observed under compressive loading along $\langle 111\rangle$ directions ($SF_{\mathrm{t}}$ ≈ 0.31 and $SF_{\mathrm{a}}$ ≈ 0.16) and $\langle 110\rangle$ directions ($SF_{\mathrm{t}}$ ≈ 0.47 and $SF_{\mathrm{a}}$ ≈ 0.24). The activation of classic twinning under conditions of markedly reduced $SF_{\mathrm{t}}$ values was previously examined in Ref. [29]. These considerations are directly relevant to uniaxial tensile experiments in $\langle 110\rangle$-oriented nanocrystals, where the onset of anti-twinning processes has been proposed. The advent of these high-energy pathways has been however ruled out via three-dimensional analyses of atomic trajectories (Section 2.2.1 and Supplementary Material 1), regardless of loading configuration.
Identification of the classic twinning pathway relies on determining the $\{1\bar{1}0\}$ plane normal to $K_1 = \{112\}$ that contains the $\langle 11\bar{1}\rangle$ atomic trajectories (in the absence of any out-of-plane displacements), referred to as the plane of shear $P$. Following Fig. 1, vector $\eta_2$ is then oriented along the marked $\langle 111\rangle$ direction in the parent crystal. Reflection of $\eta_2$ across the habit $K_1$ plane renders vector $\eta_{2,t}$ in the twinned crystal region, together with the characteristic correspondence angle of the classic twinning route, $\theta_t = 38.9°$. During twin formation, the atoms aligned along $\eta_2$ in the parent crystal are thus realigned along $\eta_{2,t}$.

**2.2. The BCC-HCP-BCC twinning pathway in Ta, Nb and Fe**

*2.2.1. Atomic shuffles driven by compression*

The results sketched in Fig. 2 illustrate the distinctive dual-shuffle pathway observed on $\{110\}$ planes of Ta, Nb and Fe nanocrystals when subjected to uniaxial compression along $\langle 100\rangle$ directions where $SF_{\mathrm{a}}$ > $SF_{\mathrm{t}}$ (i.e., $SF_{\mathrm{t}} \approx 0.24$ and $SF_{\mathrm{a}} \approx 0.47$). This pathway, summarized in Table 1, falls along the lines of the symmetry breaking analyses in Ref. [30] which correlates parent and twinned BCC configurations through saddle or transitional HCP structures. The observed HCP-mediated twinning route comprises a first pressure-induced shuffle event, incumbent to the reduction of the interatomic separation along the $\langle 100\rangle$ loading axis (Figs. 2(a)-(c)). This process is enabled by the structural instability of $\{110\}$ layers, triggered by the relocation of surface energy valleys under the critical applied compression, $\sigma_{\mathrm{c}}$. As indicated in Fig. 2(c), coordinated atomic displacements are then produced towards these nascent energy valleys, resulting in consecutive shuffles of $\{110\}$ layers along $\langle 1\bar{1}0\rangle$ and $\langle\bar{1}10\rangle$ directions. This produces the zigzagged atomic arrangement illustrated in the Fig. 2(a), along with the transition of the ABAB… stacking sequence of the parent BCC crystal into that of an HCP structure across the basal $\{0001\}$ plane. It shall be noted that the first shuffle event cannot be shear-driven because the critical resolved shear stress $\tau_{\mathrm{r}}$ vanishes on the shuffled $\{110\}$ layers under $\langle 100\rangle$ compression ($SF = 0$). The currently observed pathway supports the transformation process under uniaxial stress proposed in Ref. [31], in contrast to the Burgers shear-driven path for the BCC-to-HCP transformation in Refs. [32–34]. This shuffle process consistently yields the Pitsch-Schrader Orientation Relationship (OR) between HCP and BCC lattices [35].
An abrupt second atomic shuffle process then occurs simultaneously with lattice relaxation (expansion) along $\langle 111\rangle$ directions of the parent BCC lattice as the stress decreases during the plastic pop-in event at the constant applied displacement ($\sigma < \sigma_{\mathrm{c}}$), Fig. 2b. This process is attendant on the creation of the new energy valleys on the A-layers, sketched at the right-hand side of Fig. 2(c). Atoms in the basal B-layers of the transitional HCP structure are then shuffled along $\langle 113\rangle$ directions in the compressed parent BCC crystal (or $\langle 1\bar{1}00\rangle$ directions of the HCP crystal) so as to occupy the newly formed valleys, ultimately producing the twinned BCC atomic configuration.
The net shear strain carried by the dual-shuffle process arises from the elastic compressive strains —triggering the critical zigzag atomic configuration in the first shuffle event— and the lattice expansion which enables the second shuffle event. These two processes are highlighted with black arrows in Fig. 2(b), leading to the net $\langle 111 \rangle$ displacement indicated with the red arrow. Although comparable net atomic displacements along $\langle 111 \rangle$ directions are produced in both the dual-shuffle and classic twinning pathways, the effective atomic trajectories in the dual-shuffle mode lead to the correspondence angle $\theta_{\mathrm{ds}} = -15.8°$ which differs from the value of $\theta_{\mathrm{t}} =38.9°$ in the classic mode. Fig. 2(c) summarizes these correspondence angles together with that of the hypothetical shear-driven anti-twinning process ($\theta_{\mathrm{at}} = -41°$). Details concerning the identification of the $\{110\}$ *P*-plane are given in Supplementary Material 1, Fig. S1.

Figure 3 illustrates the twin boundary morphology produced by the dual-shuffle pathway in a variety of loading conditions. Under uniaxial compression, the non-uniform extension of the zigzagged atomic configuration on successive $\{110\}$ shuffle layers gives rise to rumpled or staggered boundaries (Fig. 3(a1)). Since these boundaries consist of the $\langle 111 \rangle$ terminations in each shuffle layer, the variable extension of these layers is not necessarily conducive to the formation of the distinctive straight $\{112\}$ $K_1$ plane of the classic twinning route (Fig. 3(a3)). The BCC-HCP-BCC twinning pathway therefore constitutes an invariant-line, rather than an invariant-plane, transformation process.

An alternative mechanism yielding the same correspondence angle as the dual-shuffle process has been proposed in the form of a twinning/anti-twinning pathway operating in $\{112\}$ planes along $\langle 111 \rangle$ directions under the dominant action of shear stresses [18,19]. Although both mechanisms produce fundamentally distinct atomic trajectories, the same correspondence angle $\theta_{\mathrm{ds}}$ and net shear strains along the anti-twinning sense arise. By contrast, while the postulated twinning/anti-twinning pathway generates net shear strain via partial dislocation glide on $\{112\}$ planes, the pressure-driven dual-shuffle mechanism produces plastic deformation through coupled lattice compression and expansion processes. The characteristic twinning symmetry of the BCC structure therefore arises from the lattice reorientation produced by such coupled compression and expansion processes, rather than from the shear localized on a dislocation glide plane.

### *2.2.2. Twin morphologies under different loading conditions*

Figures 3(a1) and (a2) showcase the distinctive wedge-shaped twins nucleated at the nanocrystal surface during uniaxial compression. These twins are composed of shuffled $\{110\}$ layers (or $P$-planes), bounded by the staggered surfaces discussed in Section 2.2.1. Twin heightening proceeds through the extension of zig-zagged atomic chains in the *h* direction, whereas frontal twin growth is achieved by incorporating zigzagged chains along the *f*-direction (Fig. 3(a2)). Wedge widening occurs concurrently through the addition of successive sets of shuffled $\{110\}$ layers along the *w*-direction. Surface protrusions also arise due to the accommodation of the shear strains in twin variants (Fig. 3(a2)).

Since the same rotational symmetry of BCC twins must be achieved regardless of the active twinning pathway, the atomic positions reached via the classic shear-driven mechanism necessarily coincide with those produced through the dual-shuffle pathway. This equivalence enables the formation of faceted twin units in which straight $\{112\}$ boundaries coexist with staggered boundaries on intercepting facets, as observed in nanocrystals compressed within ≈20° misorientations from the $\langle 100 \rangle$ loading axis (see Fig. 3(a4)). These mixed defect structures comprise $\{112\}$ twin facets developing through the shear-driven classic pathway and $\{110\}$ facets on which the pressure-driven dual-shuffle process arises.

Our simulations also show incipient dual-shuffle processes in $\langle 110 \rangle$ and $\langle 111 \rangle$-oriented nanocrystals subjected to tensile loads (Fig. 3(c)). Here, the formation of the transitional HCP phase is facilitated by the development of new surface energy valleys as the A-layers are stretched along the *t*-direction indicated in Fig. 2(c). The incipient shuffled planes produced under tension rapidly evolve into dislocation loops, as illustrated in Fig. 3(c), demonstrating that the dual shuffle pathway can act as a precursor to archetypal dislocation nucleation processes.

Figure 3(b) shows the twin structures emerging in nanoindentation simulations of $(100)$ surfaces. Plastic deformation initiates through the nucleation and growth of lenticular twins [36–38] (Fig. 3(b1)) which form through the cooperative pairing of the dual-shuffle pathway and the classical twinning route. These collective atomic processes hinge on the capacity of the dual-shuffle mechanism to enable coordinated extensions of the shuffled $\{110\}$ layers (Fig. 3(a3)), which facilitates the development of habit $\{112\}$ boundaries essential to any subsequent shear-driven twin propagation. Defect analyses beneath at the nanoindentation core region further indicate the predominance of the dual-shuffle pathway under the extreme compressive stresses generated during contact (Fig. 3(b2)). These observations highlight the pervasive role of this pathway in the formation of the intertwined twin network (or crystallites) reported in [37], consisting of a myriad of shuffled $\{110\}$ layers.

Although the formation of wavy twin boundaries has been experimentally observed in BCC metals at *micrometer* scales, this feature has been attributed to the activation of the classic shear-driven twinning mode. These wavy boundaries fundamentally differ from the atomic scale staggered configurations arising under the dual-shuffle pathway (see Supplementary Material 2).

## 2.3. The BCC-FCC-BCC twinning pathway in W and Mo

### *2.3.1. Shear-driven BCC to FCC-like transitions under extreme tension and compression*

Building on the crystallographic analyses in Ref. [30], we identify a twinning pathway mediated by the emergence of transitional FCC-like phases in our simulations of uniaxial compression and crack-tip deformation processes in $\langle 100\rangle$-oriented W and Mo nanocrystals (Fig. 4 and Table 1). This mechanism prevails in BCC nanocrystals with higher elastic stiffness, thereby superseding the HCP-mediated pathway that dominates in the elastically softer Ta, Fe and Nb nanocrystals. The initial stage of the twinning pathway in W and Mo involves a Nishiwama-Wasserman (N-W) transformation, in which a derivative FCC-like structure develops via two coupled processes (see Fig. 4(a1) and Refs. [39,40]): (i) a Bain-type lattice reorientation, involving a $\approx$45° axis-rotation process which produces a quasi-FCC, tetragonal structure; and (ii) a structural rotation by the angle $\delta$ ranging from $\approx$7° (uniaxial compression) to $\approx$10° (crack-tip deformation). The resulting FCC-like structure can be rationalized in terms of partial dislocation glide, occurring either within the parent BCC structure or the quasi-FCC structure generated upon Bain reorientation (Fig. 4(a2)). In the former case, the partial dislocations sweep on $\{110\}_{\mathrm{BCC}}$ planes with a net $b = (a/6)\langle 110\rangle$ —corresponding to the complementary vectors $b_1$ and $b_2$ — whereas in the latter case, the partial dislocations glide on consecutive $\{111\}_{\mathrm{FCC}}$ planes with $b = (a/6)\langle 112\rangle$. Notably, the lattice structures supporting these dislocation processes become distorted by the imposed elastic strain fields during uniaxial compression or crack-tip deformation, directly affecting on the measured $\delta$. The resulting N-W orientation relationship (OR) then satisfices $(111)_{\mathrm{FCC}} \| (110)_{\mathrm{BCC}}$; $[\bar{1}10]_{\mathrm{FCC}} \| [001]_{\mathrm{BCC}}$; $[112]_{\mathrm{FCC}} \| [110]_{\mathrm{BCC}}$ [22], as evidenced by the BCC/FCC interfaces in Fig. 4(b).

The structural similarity between the ideal FCC lattice and the derivative FCC-like structures produced through the N-W pathway is supported by polyhedral template matching (PTM) analyses, which enable robust identifications of crystal structures under large stresses and thermal fluctuations [41]. Under uniaxial compression, the Root Mean Square Deviation (RMSD) of the FCC-like lattices reaches $\approx$0.13. Although this value departs from the ideal RMSD = 0 of the perfect FCC structure, it remains well below the limiting level of $\approx$0.20 at which FCC and BCC structures become equivalent or indistinguishable. In the simulations of crack-tip plasticity, the FCC-like phase also deviates from the perfect FCC lattice; however, the structural resemblance is further enhanced, as reflected by a reduced RMSD of approximately 0.08. A key factor contributing to the cuboidal character of the FCC-like structures involves the atomic displacements resulting from the applied uniaxial compression, or the tensile stress prevailing at the vicinity of crack-tips, as both loading conditions reduce the tetragonality of the final structure. Following the representation in Fig. 4(a1), the cuboidal shape of the FCC-like structure is enhanced when biaxial compression (rather than uniaxial compression) is applied along the two equal, larger lattice parameters $a_x = a_y$. Similar results are attained when tensile stresses prevailing at crack tips increase the smaller lattice parameter $a_z$.

Figure 5 illustrates the slender, platelet-shaped structure generated by the transient FCC-like phase. These platelets propagate from the surface in uniaxially compressed nanocrystals (Fig. 5(a1)) or from the vicinity of crack tips in nanocrystals loaded in tension (Fig. 5(b1)). Moreover, when uniaxial tensile loadings are applied along $\langle 111 \rangle$ directions, platelet nucleation and growth is triggered by the resolved shear stress acting on critical $\{110\}\langle 111 \rangle$ systems, leading to intersecting platelet arrays (Fig. 5(a4)). The FCC-like platelets exhibit coherent, stepped interfaces in which $\{110\}$ planes of the surrounding BCC region align with $\{111\}$ planes of the inner FCC-like region, as sketched in Fig. 4(b). Consistent with the N-W OR, interface formation is mediated edge dislocations, or disconnections, with $b = (a_{\mathrm{FCC}}/6)\langle 112 \rangle$, giving rise to regularly stepped interfaces (Fig. 4(b)).

*2.3.2. Emergence of the twinned BCC phase under variable loading conditions*

Three essential atomic pathways are observed by which the transient FCC-like phase transforms into the twinned BCC structure in $\langle 100 \rangle$-oriented W and Mo nanocrystals subjected to either uniaxial compression (Fig. 5(a)) or crack tip deformation (Fig. 5(b)), as summarized in Table 1.

In uniaxial compression a composite platelet structure traversing across the nanocrystal is formed (Fig. 5(a1)). Platelet nucleation involves the formation of incipient FCC-like layers at the nanocrystal surface, accompanied by the activation of the dual-shuffle route within the inner BCC region (Figs. 5(a2) on which the correspondence angle $\theta_{ds} = -15.8°$ is measured. This layered structure subsequently propagates toward the nanocrystal interior, with the twinned BCC interlayer trailing the HCP phase at the forefront. The shear strain generated by dislocation glide on the $\{111\}$ planes of the FCC layers plays a decisive role in promoting surface bulging and in facilitating the activation of the dual-shuffle pathway on the inner BCC layer. Consistent behavior is observed in nanoindentation simulations of $(100)$-oriented W and Mo surfaces, which depict the formation of incipient FCC-like platelets on inclined sets of $\{110\}$ planes beneath the contact surface (Fig. 5(c)). This process is again followed by the formation of staggered twin units in adjacent regions through the dual-shuffle pathway.

A different scenario of atomic displacements is observed away from the free surface in uniaxially compressed freestanding nanocrystals, or in bulk crystals in the absence of free surfaces (Supplementary Material 3, Fig. S2). In freestanding nanocrystals, FCC bands that propagated from the surface subsequently become annihilated (Fig. 5(a3)). This process produces defect clusters with substantial out-of-plane displacements along with regions with the parent BCC configuration on which the dual shuffle pathway ultimately activates, enabling the formation of twinned BCC pockets (Fig. 5(a3)).

In the vicinity of crack tips, the nucleation of FCC-like bands follows a similar pathway as that described in Section 2.3.1, involving a first N-W transformation which carries the lattice rotation of $\delta_1 \approx 10°$, and a 45° axis-reorientation. A second N-W transformation subsequently develops within the FCC-like bands, characterized by an additional 45° axis-reorientation and a further lattice rotation, $\delta_2 = \delta_1 \approx 10°$ (Fig. 5(b1)). Since these two N-W transformations occur on different $\{110\}$ planes, the twinned region within the FCC-like bands and the surrounding parent BCC crystal exhibit a clear misorientation with respect to the conventional twinning symmetry. Nevertheless, distinctive $\{112\}$ twin boundaries ultimately arise as the twinned region propagates across the supporting FCC platelet, impinging on the surrounding parent BCC crystal (Fig. 5(b2)).
A discussion is given in Supplementary Material 2 concerning the experimentally observed BCC-FCC-BCC transformation pathway developing ahead of crack tips.

## 2.4. Energy landscapes for twin nucleation and critical pop-in stresses

*2.4.1. The classic twinning pathway*

Figure 6(a) summarizes the energy landscapes for the classic twinning pathway. The fault energy ($\gamma$)—shear displacement ($s$) curves support the notion that when a twin nucleus consisting of a minimum of $n = 3$ layers (one $\{112\}$ interlayer and two $\{112\}$ boundaries) reaches a peak energy barrier $\gamma_{\mathrm{max}}$, spontaneous relaxation occurs towards the saddle point $\gamma_{\mathrm{min}}$ located at the value of $s = (a/6)\ \langle 111\rangle$ which underlies twin formation (Figs. 6a(1) and (a2)).
The results in Figs. 6(a3) and (a4) indicate that the application of either uniaxial compression along $\langle 100\rangle$ directions or hydrostatic pressure increases the system energy by reducing the interatomic separation between adjacent $\{112\}$ layers. This effect shifts the $\gamma - s$ curves to higher energy values, thereby suppressing twin nucleation. It is found that these pressure-sensitive effects tend to become more pronounced with tabGAP potentials than with EAM potentials, while the increase in $\gamma_{\mathrm{max}}$ under *uniaxial compression* also exceeds that attained under *hydrostatic pressure*.
Figure 7 presents uniaxial stress—strain curves along different crystallographic directions. These results indicate that the twofold greater $\gamma_{\mathrm{max}}$ reached in W compared to in Ta under the applied tension (cf. Figs. 6(a3) and (a4)), produces a critical pop-in stress $\sigma_{\mathrm{c}}$ in W that is also approximately doubled (see Fig. 7(a1)). Moreover, the value of $\gamma_{\mathrm{max}}$ setting twin initiation is fundamentally affected by non-linear elastic responses arising at high stresses, where the interatomic distances fall out of the range of the harmonic approximation. Therefore, the progressive reduction in the Young's modulus $E = d\sigma/d\varepsilon$ (elastic softening) observed under tension along $\langle 100\rangle$ and compression along $\langle 110\rangle$ orientations (Fig. 7(a)) marks the onset of twinning instabilities, which limit $\gamma_{\mathrm{max}}$ and $\sigma_{\mathrm{c}}$. The reversed trend is observed along $\langle 111\rangle$ orientations (Fig. 7(a3)) where $E$ is progressively increased (elastic hardening).
Our analyses reveal pronounced pressure effects on the critical shear stress for twinning $\tau_{\mathrm{c}}$, consistent with the observation of higher $\gamma_{\mathrm{max}}$ at increasing compression. The overall trends for different potentials are depicted in Fig. 7(b), where $\tau_{\mathrm{c}}$ is plotted against the normal stress $\sigma_{\mathrm{r}}$ (both resolved on the active $\{112\}\ \langle 11\bar{1}\rangle$ twinning systems) for $\langle 100\rangle$, $\langle 110\rangle$ and $\langle 111\rangle$-oriented nanocrystals. The figure showcases a clear saturation of pressure sensitive responses when the normal stress on the twinning systems $\sigma_{\mathrm{r}}$ is reduced below ≈8 GPa.

*2.4.2. The HCP- and FCC-mediated pathways*

Figure 6(b) shows the energy landscapes dictating the onset of the first shuffle stage in the HCP-mediated twinning route. The trends observed for an infinite array of shuffled $\{110\}$ layers (Figs. 6(b1) to (b4)) reveal a sustained reduction in the slope $\mathrm{d}\gamma/\mathrm{d}s$ during the initiation of the shuffle process, a feature that hinges on the application of high pressures. The spontaneous nature of the first shuffle event becomes evident since for a critical compressive stress, it ultimately follows that $\gamma_{\mathrm{max}} \rightarrow 0$ and $\mathrm{d}\gamma/\mathrm{d}s < 0$. Critical zigzagged atomic configuration then naturally unfolds as the system relaxes towards $\gamma_{\mathrm{min}}$ where the HCP-like structure forms.

While the classic twinning pathway is progressively hindered under compression due to the increase of the energy barrier $\gamma_{\max}$, the dual-shuffle route is activated once the critical compressive stress is reached along $\langle 100 \rangle$ directions. This feature is central in understanding nanoindentation responses in $(100)$ surfaces, where in spite of the potential activation of multiple twinning systems in the subsurface under the resolved shear stresses, the dual-shuffle process activates first owing to the high applied contact pressure.
The $\gamma - s$ curves further reveal the inherently imperfect nature of the transitional HCP phase. While a perfect HCP structure requires a precise combination of compressive strain and shuffle distance $s$, the first shuffle event produces imperfect HCP lattices which relax at $\gamma_{\min}$ under variable $s$ values (Fig. 6(b)), resulting in deviations from the ideal ratio of $c/a$ =1.63 between lattice parameters.
Our results also show that the $\gamma - s$ curves shift to increasing energy levels under conditions of elevated hydrostatic pressure ($\nu = 0$), eventually suppressing the spontaneous nature of the dual-shuffle pathway (Figs. 6(b1) and (b2)). Moreover, comparison between simulations for infinite $\{110\}$ layers and those for a single shuffled layer (cf. Figs. 6(b1)—(b4) and 6(c)) indicates that the energy minimum $\gamma_{\min}$ vanishes in the latter case. This suggests that the spontaneous character of the dual-shuffle pathway at high compression requires the onset of supercritical twin nuclei comprising a minimum number of $\{110\}$ layers so as to ensure that $\mathrm{d}\gamma/\mathrm{d}s < 0$.
Finally, our simulations show that the energy landscapes for the onset of the FCC-mediated twinning pathway remain monotonically increasing, without distinct $\gamma_{\max}$ or $\gamma_{\min}$ levels (Fig. 6(d)). Nevertheless, as the applied compression sharply increases and the stress—strain curves gradually soften ($E = d\sigma/d\varepsilon < 0$), the energy landscapes develop increasing concavity ($\mathrm{d}^2\gamma/\mathrm{d}s^2 < 0$) near $s \approx 0.25a_{\mathrm{BCC}}$, coinciding with the slip distance carried by the dislocation glide process on $\{110\}$ planes which leads to the formation of the FCC phase.

### 2.5. The impact of stable phase transitions upon twin formation

Our results in Fig. 8 provide a thermodynamic perspective on twin nucleation mediated by saddle HCP structures, extending earlier investigations in Refs. [20,21,25,42–44]. While the transitional HCP phase produced in the compressed Fe nanocrystals corresponds to the thermodynamically stable high-pressure $\varepsilon$-Fe phase (Figs. 8(a2)—(a4)), the transitional HCP phase generated in Ta nanocrystals (Fig. 8(a1)) differs from the stable hexagonal $\omega$-Ta phase, which only forms at substantially higher hydrostatic pressures [45,46]. This difference has profound consequences for the ensuing twinning processes.
Although the second shuffle event into the twinned configuration proceeds almost instantly upon the nucleation of the *metastable* HCP phase at the pop-in stress ($\sigma = \sigma_{\mathrm{c}}$) in Ta nanocrystals, the dual-shuffle pathway in Fe nanocrystals is attendant on the development of the thermodynamically stable $\varepsilon$ phase which progressively transitions into the twinned BCC structure only when the applied compression is significantly reduced during the pop-in event ($\sigma \ll \sigma_{\mathrm{c}}$). The stability of the transitional HCP $\varepsilon$-Fe phase over a wider range of applied stress enables the occurrence of basal-plane rotations [21,47], as evidenced by the angle $\alpha_{\mathrm{HCP}} \approx 5°$ (Fig. 8(a2) and (b2)). These rotations generate substantial shear strain localization and surface tilts, either through the propagation of HCP regions nucleated on free surfaces or through the growth and coalescence of dispersed HCP domains in a bulk crystal (Ref. [48]). As the second shuffle invariably initiates at the interior of rotated HCP regions in Fe, the surface tilts exhibit a distinctive angle $\alpha_{\mathrm{ds}}$ (Fig. 8(a4)).
During the propagation of the twinned BCC phase across the HCP $\varepsilon$-Fe platelets, low-energy (relaxed) $\{112\}$ boundaries develop with the parent BCC region (Fig. 8(a4)). These straight interphases clearly contrast with the staggered boundaries that form at $\sigma \to \sigma_{\mathrm{c}}$ under the high elastic energy stored in Ta nanocrystals. In this case, the second shuffle process proceeds almost instantly at the interior of *unrotated* domains in a bulk crystal (Fig. 8(b3) and Supplementary Material 4) or at the forefront of surface nucleated twin platelets (Fig. 3(a2)), in the absence of shear band formation.
A discussion on experimental findings of twinning processes mediated by the formation of stable HCP and FCC phases is given in Supplementary Material 2 and 4 (see Figs. S3 and S4).

## 3. Concluding remarks

This investigation establishes a new paradigm for deformation twinning in metallic BCC nanocrystals by revealing intricate, transformation-mediated atomic pathways that operate at spaciotemporal scales beyond current experimental resolution. By resolving the transient formation of HCP- and FCC-like phases as a function of elastic stiffness, we provide a unified mechanistic framework for the onset of nanoscale plasticity and the emergence of distinctive defect structures associated with competing twinning pathways. This knowledge connects prior point-group symmetry analyses in [30] with the general defect processes, atomic pathways, and critical stresses associated with the emergence of transformation-mediated twinning in small length scales across a broad range of loading conditions. Our findings provide a fundamental framework to the analysis of flow-stress asymmetries in BCC nanocrystals [13–16] by identifying twinning mechanisms that activate at distinct stress levels depending on loading direction, while ruling out the emergence of recently proposed shear-driven anti-twinning modes [17]. The consistency of the identified atomic scale processes across multiple BCC metals modeled under a variety of interatomic potentials—including quantum-level accurate tabGAP and EAM potentials—indicates that transformation-mediated twinning is a general feature in freestanding BCC nanocrystals, as well as in bulk crystals where free surfaces and preexisting defects are absent.

The following are the salient results from our investigation:

**1.** In Ta, Nb and Fe nanocrystals of moderate elastic stiffness, we identify a pressure-driven lattice instability in which the parent BCC structure transitions into HCP lattices, followed by the formation of the twinned BCC configuration via a dual-shuffle pathway. In marked contrast to classic twinning—governed by shear-driven glide of partial dislocations on $\{112\}$ habit planes—this HCP-mediated dual-shuffle pathway unfolds through a fundamentally distinct scenario of atomic displacements on $\{110\}$ planes, which produces intrinsically staggered twin boundaries. The permanent shear strains originating from this twinning route involve specific lattice expansion and relaxation processes along individual $\langle 111\rangle$ lines in the absence of characteristic $\{112\}$ planes, challenging the fundamental hypothesis in metal plasticity where shear strains arise from the action of the resolved shear stresses on characteristic planes where dislocations typically glide.
A central outcome from our simulations is the formation of invariant $\langle 111\rangle$ symmetry lines which separate parent from twinned crystal regions. The mobilization of these $\langle 111\rangle$ lines, or surface steps, allows for the development of arbitrary twin boundaries, including conventional low-energy $\{112\}$ surfaces. As the formation of such $\{112\}$ boundaries ultimately enables twin growth to proceed through the distinctive partial dislocation glide process of the classic twinning route, a synergistic interplay develops between the twinning modes.
We demonstrate that the twin structures that emerge from the dual-shuffle pathway are fundamentally governed by the stability of the transient HCP phase. In Ta nanocrystals, where the parent BCC crystal transforms into a metastable HCP phase, high-energy staggered boundaries develop on distinctive wedge-shaped twins and platelets. By contrast, in Fe nanocrystals, the formation of a stable HCP phase promotes collective lattice rotations about the basal plane, leading to pronounced shear banding and the emergence of low-energy $\{112\}$ boundaries. This process underlies formation of the laminated twin structures observed in Fe nanocrystals [48,49], which thicken under continued loading [33,44,49–52]. Consistent simulation results are observed in bulk BCC crystals lacking free surfaces, which reveal pronounced shear banding in Fe and its absence in Ta.

**2.** Our simulations reveal the prevalence of an ultrafast twinning route mediated by the formation of unstable FCC-like platelets in the elastically stiffer W and Mo nanocrystals. We establish the atomic pathway to the transition from the parent BCC lattice to a highly distorted FCC-like phase by resolving the details of its imperfect structure along with the onset of a Nishiwama-Wasserman (N-W) orientation variant, and the formation of highly-mobile semi-coherent BCC-FCC interphases. In marked contrast to the HCP-mediated twinning route which is driven by pressure, the formation of the transient FCC phase is governed by the action of large resolved shear stresses.

We identify multiple atomic pathways by which the transient FCC-like phase transforms into the twinned BCC configuration. In the vicinity of crack tips, a second N-W transformation nucleates and propagates within the FCC-like platelets, ultimately leading to the formation of twin bands with conventional $\{112\}$ twin boundaries. At the surfaces of nanocrystals subjected to uniaxial compression or nanocontact loading, the shear strain carried by distorted FCC-like layers promotes activation of the HCP-mediated dual-shuffle pathway in adjacent regions, resulting in staggered twin boundaries. Finally, as the FCC-like platelets nucleated at the surface progressively annihilate at the nanocrystal interior, localized twin pockets emerge together with regions of high defect density.

**3.** The energy landscapes governing the transformation from the parent BCC phase into transitional FCC- and HCP-like phases offer fundamental insights into the competition between twinning pathways depending on the elastic stiffness of the metallic material. Our results demonstrate that the classical twinning route is progressively suppressed with increasing uniaxial compression or hydrostatic pressure due to a systematic increase in the energy barrier $\gamma_{\max}$ that must be overcome for its activation, as revealed by the fault energy ($\gamma$)—displacement ($s$) curves. This feature leads to resolved twinning stresses that scale proportionally with pressure in all BCC metals.

In BCC metals of moderate elastic stiffness under high compressive stresses, the plateauing of the $\gamma(s)$ curves in the vicinity of $s = 0$ where $d\gamma/ds < 0$, promotes the spontaneous formation of supercritical nuclei composed of multiple $\{110\}$ planes on which the HCP-mediated dual shuffle process activates. This pressure-driven plastic instability arises independently of any resolved shear stress. At even higher compressive stresses in elastically stiffer BCC crystals, a decrease in the curvature of the energy landscape ($d^2\gamma/\mathrm{d}s^2$) occurs for $s \approx 0.25a$, facilitating the partial dislocation glide process on $\{110\}$ planes associated with the onset of the FCC-mediated twinning pathway. In contrast to the HCP-mediated route, the inception of the transitional FCC phase is highly unstable, as evidenced by the absence of energy minima in the $\gamma(s)$ curves. The structural instability of the FCC phase then leads to its rapid annihilation along with the formation of the stable twinned BCC structure.

## Methods

### MD and DFT simulations

This investigation comprises a comprehensive set of large-scale MD simulations of nanocrystal compression, tension, and nanoindentation, and crack tip plasticity performed with the LAMMPS code [53]. The simulations concern archetypal group V, VI and VIII BCC transition metals modelled with the embedded-atom method (EAM) potentials in [54] for Ta, [55] for W, and [56] for Fe; the angular-dependent potential (ADP) in [57] for Fe; and the machined-learned (tabGAP) quantum-level accurate potentials for Ta, W, and Nb trained with density functional theory (DFT) simulations [58]. The accuracy of the different potentials in capturing pressure and shear-driven twinning pathways is discussed in Supplementary Material 5 (Figs. S5 to S7). Complementary simulations were performed with BCC cells subjected to periodic boundary conditions to investigate twin nucleation processes in bulk crystals or superlattices. A complete set of brute-force DFT simulations was also conducted for the assessment of the energy landscapes associated with the different twinning pathways. These simulations employed small unit cells and were carried out with the VASP code [59], where the PBE exchange-correlation functional was used [60] along with PAW potentials that include semi-core electrons [61]. The plane-wave energy cutoff was 500 eV and the k-point grid was automatically created using a maximum spacing of 0.15 $\text{Å}^{-1}$.
Cylindrical computational cells were subjected to uniaxial compression or tension to simulate the stress-strain curves in nanocrystals with 8 nm in diameter and 18 nm in height. The canonical NVT ensemble was invoked in all simulations used thermalized MD cells. The cells were subjected to strain rates ranging from 0.1 $ns^{-1}$ to 0.01 $ns^{-1}$ under displacement control along $[100]$, $[110]$, and $[111]$ orientations, including a number of orientations $20°$ away from $[100]$, with time step of 1 fs. Periodic boundary conditions were applied to the top and bottom surfaces during straining. Simulations for *bulk* crystals where free surfaces and preexisting defects are absent were also performed by subjecting fully-periodic cuboid computational cells of 13 × 13 × 13 $nm^3$ to uniaxial tension or compression. The NPT ensemble was invoked in these simulations while the imposed deformation rates ranged from 0.1 $ns^{-1}$ to 0.01 $ns^{-1}$ (using a timestep of 1 fs). Vanishing pressure was maintained in the lateral directions.
The role of the applied loading rate upon defect nucleation and growth was not significant, even though the processes by which multiple nanodomains coalesce into individual shear bands in the uniaxial compression simulations of Fe supercells, including those that lead to the formation of FCC platelets around crack tips, were less noticeable at higher strain rates. A few complementary MD simulations were also conducted at 0.001 $ns^{-1}$ to ensure consistency of defect structures.
The nanoindentation simulations were carried out using cuboidal-shaped MD cells of $70 \times 70 \times 40$ $nm^3$ brought into contact with a rigid spherical tip of diameter $D = 24$ nm, modeled through a repulsive potential [37]. The simulations were performed under the canonical NVT ensemble using the Nosé-Hoover thermostat, a timestep of 2 fs, and an indenter tip velocity $v = 2$ m/s. The indenter tip penetration process was modelled on $\{100\}$, $\{110\}$ and $\{111\}$ surfaces. Periodic boundary conditions were applied on the lateral sides of the MD cells, while the atoms at the bottom surface remained fixed. In spite of the large applied indenter tip velocity, extensive MD analyses in BCC metals have shown that the variability of the pop-in load becomes smaller than 5% as $v$ is further reduced to 0.1 m/s, and is essentially attributable to statistical variations of the MD ensemble during multiple realizations [37].

Simulations of crack tip plasticity were performed using rectangular MD cells of $x = 10$ nm, $y = 40$ nm, and $z = 60$ nm that contained a central slit crack of 7 nm in length, oriented along the $x$-axis. The cells were loaded along the $y$-axis in displacement control with velocity of 2 m/s under the NVT ensemble (timestep of 2 fs). Periodic boundary conditions were imposed on the three principals $x$-, $y$-, and $z$-axes.

Prior to the nanoindentation and crack tip plasticity simulations, all MD cells were subjected to a thermalization process to minimize the internal energy [37].

**Simulations of energy landscapes**

The energy landscapes for the onset of the different twinning modes were evaluated using the LAMMPS code [53]. The nucleation of twins containing $n$ $\{112\}$ layers under the classic mode was carried out following the methods detailed in [29]. In short, we constructed an orthogonal BCC cell with crystal directions $[110]$ along $x$, $[112]$ along $y$ and $[111]$ along $z$, with dimensions 4 × 20 × 18 atomic layers (1440 atoms). The system was relaxed to zero pressure after which a fixed set of $n$ $\{112\}$ central layers were sequentially displaced in the z-direction through the application of a shear displacement $n \times s$ (where $s$ is the incremental slip magnitude by which the adjacent layers are displaced). The interplanar distances were optimized by enabling atomic relaxations in the $y$ direction. The shearing process requires non-periodic boundaries in $y$, while a number of top and bottom layers remained fixed to avoid having spurious surface relaxation effects. The effect of pressure was investigated by pre-relaxing the system to different levels of hydrostatic pressure as well as of uniaxial compression on the subsequently sheared $\langle 112 \rangle$ directions. The fault energy was calculated as the energy difference between the strained and unstrained computational cells divided by the shear plane area.

To explore the energy landscapes for the initiation of the HCP-mediated twinning route, a shuffling process of $(1\bar{1}0)$ planes in $[110]$ directions was modelled either with or without compressive strains being applied in the perpendicular $[001]$ direction. This was done with BCC cells oriented with $[001]$ along $x$, $[110]$ along $y$, and $[1\bar{1}0]$ along $z$. When applying compressive strains in $[001]$, the associated crystal expansion in $y$ and $z$ was imposed according to the Poisson ratio of the crystal, obtained from an independent set of simulations with different interatomic potentials or through DFT simulations. Two modelling setups were employed. First, a shuffle distance $s$ was applied between two adjacent $(110)$ layers, producing a single interface or fault plane at the boundary of the periodic cell. Secondly, an increasing shear magnitude $s$ was applied *alternately* in the forward and reversed directions on an infinite array of $(110)$ layers using a fully periodic computational cell. While the first setup involved the use of an elongated cell comprising 20 layers to eliminate self-interactions from the shuffled interface with its periodic image, the second setup involved a small periodic cell of only 4 atoms. Since atomic relaxation does not naturally occur along the plane normal in the single interface setup, interplanar relaxation had to be imposed.

Finally, the energy landscape for the initiation of the FCC-mediated twinning route was assessed by considering the atomic processes responsible for the formation of monatomic-thick FCC platelets. The simulations involved the application of a shear displacement $s$ on a $(110)$ layer of the parent

BCC phase along the $[1\bar{1}0]$ dislocation gliding direction, with compressive strain being applied in the $[100]$ direction. This simulation setup naturally accounted for the correct Poisson relaxation.

**Acknowledgements:** J.A. acknowledges financial support by Ministerio de Ciencia e Innovación (Grant PID2023-150171NB-I00). The work by J.O. was supported by the project MEBIOSYS (No. CZ.02.01.01/00/22_008/0004634) within the Programme Johannes Amos Comenius. J.B. acknowledges funding from the Research council of Finland through the OCRAMLIP project, grant number 354234. Grants for computer capacity from CSC-IT Center for Science are further acknowledged. J.D. acknowledges support by the European Union Horizon 2020 research and innovation program under Grant Agreement No. 857470, and from the European Regional Development Fund under the program of the Foundation for Polish Science International Research Agenda PLUS, Grant No. MAB PLUS/2018/8. Funding from the initiative of the Ministry of Science and Higher Education "Support for the activities of Centers of Excellence established in Poland under the Horizon 2020 program" under Agreement No. MEiN/2023/DIR/3795 is also acknowledged.

**Autor contributions:** J.O and J.A. designed the research and discussed the results with all authors. J.O, J.D and G.W conducted MD simulations. J.M. performed simulations and analyses of the energy landscapes. J.A wrote the manuscript.

**Competing interests:** The authors declare no competing financial or non-financial interests.

**Data availability:** All data supporting this investigation are available within the paper and Supplementary Information. Source data are available from the corresponding author upon sensible request.

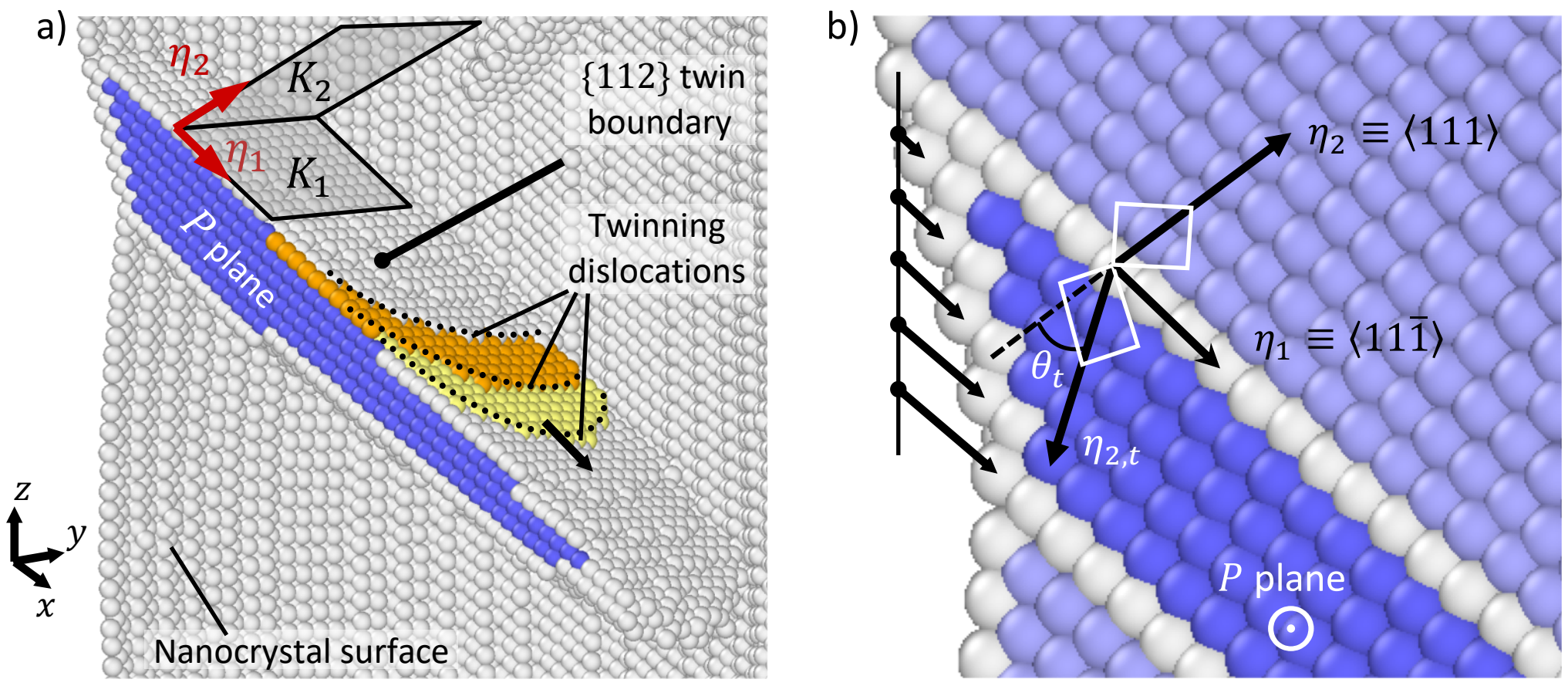

**Figure 1:** Classic twinning pathway (simulations for Ta nanocrystals under taGAP, where tensile loads are applied along the vertical <001> direction). (a) Cross-sectional view of a twin unit (where the twinned region is highlighted in blue), which thickens through the glide of the partial twinning dislocations with $b = a/6$ <$11\bar{1}$> directed along vector $\eta_1 =$ <$11\bar{1}$> that lies on the twin boundary $\{112\}$ $K_1$-plane. This process produces stepped twin boundaries, showcased in orange and yellow. (b) Detail of the twin unit in (a), indicating the $\{110\}$ plane of shear $P$ and atomic displacements produced by sequential glide of twinning dislocations from the free surface (marked with arrows). The figure highlights the vector $\eta_2$ lying within the conjugate $K_2$-plane on the parent crystal region, the complementary vector $\eta_{2,t}$ in the twinned crystal region, and the correspondence angle $\theta_t$ between the parent and twinned lattices.

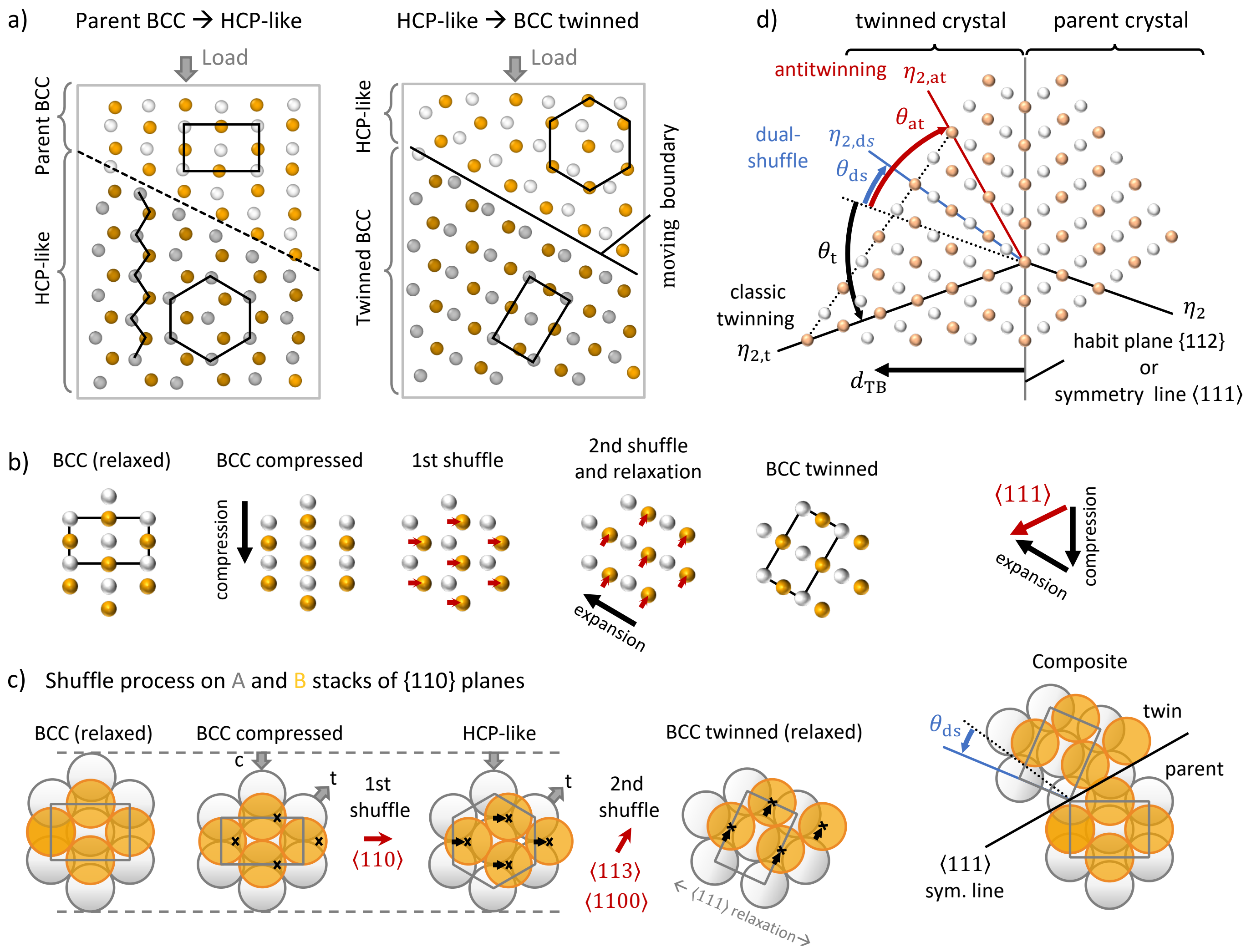

**Figure 2:** Atomic pathways for the HCP-mediated, dual-shuffle twinning route. (a) Atomic processes on the {110} shuffle planes from our Ta simulations. Notice the zigzagged atomic displacements which lead to the formation of the basal HCP plane (left-hand side), followed by the second shuffle event from the the HCP into the twinned BCC configurations (right-hand side). (b) and (c) Evolutions of the atomic arrangement in the AB stacking of the shuffled {110} planes under compression "c" or tension "t". The locations of the energy valleys in the A-stack (grey) conducive to the first and second shuffle processes are highlighted with "x". Notice the angle $\theta_{ds}$ arising during crystal relaxation in the composite view of parent and twinned regions across the ⟨111⟩ symmetry line. (c) Correspondence angles for the dual shuffle pathway ($\theta_{ds} = -15.8°$), the classic route ($\theta_t = 38.9°$) and the conceptual anti-twinning mode ($\theta_{at} = -41°$). Notice the reorientation of the vector $\eta_2$ in the parent crystal towards the corresponding directions in the dual-shuffle pathway ($\eta_{2,ds}$), the classic route ($\eta_{2,t}$), and the conceptual anti-twinning mode ($\eta_{2,at}$). Note that the same rotational symmetry between the parent and twinned lattices arises irrespective of the active twinning pathway. Moreover, the dual-shuffle pathway leads to the same incremental atomic displacements at a distance from the twin boundary, $d_{TB}$, as those produced by a twinning/anti-twinning process on {112} planes (see text for details).

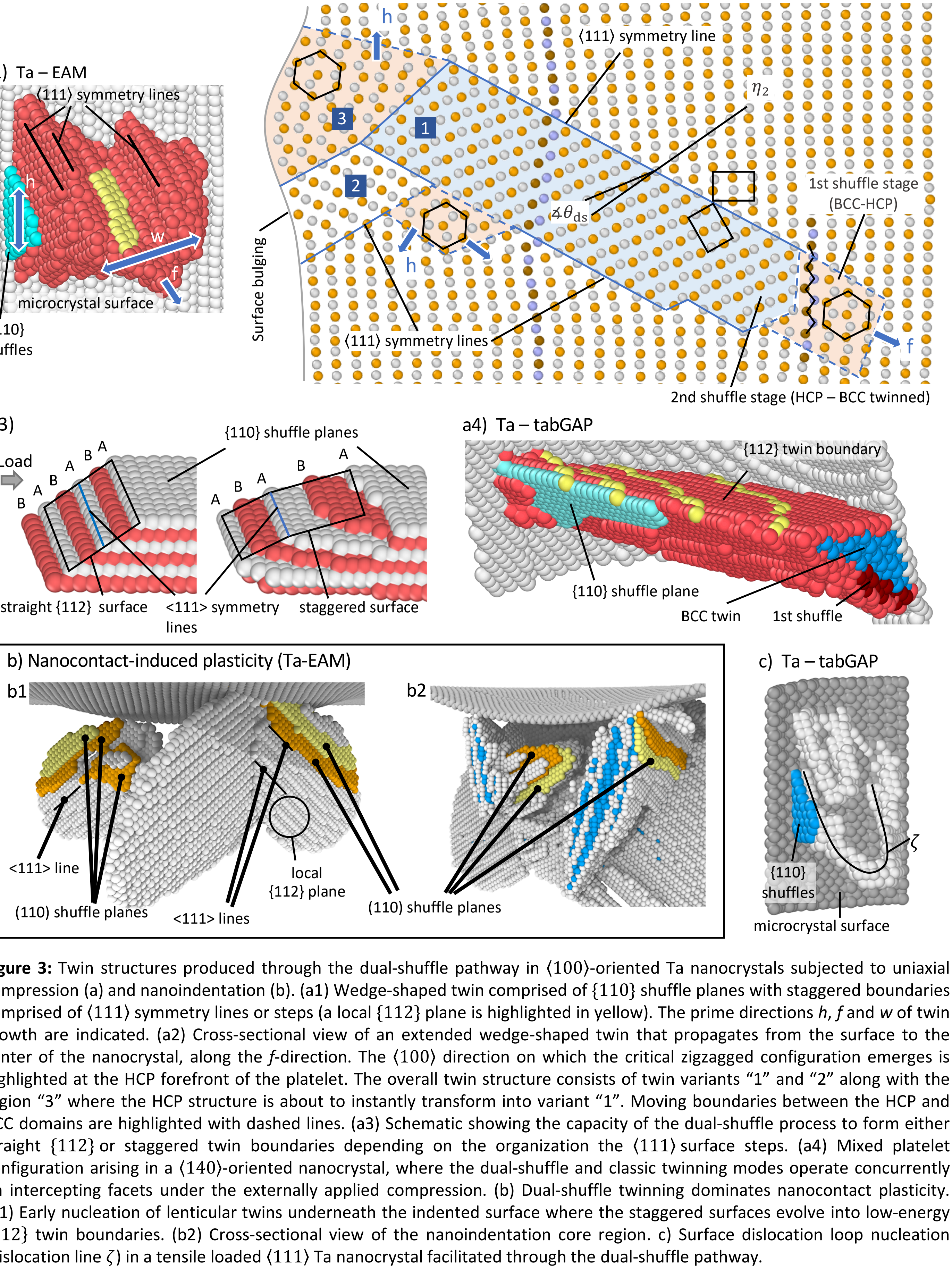

**Figure 3:** Twin structures produced through the dual-shuffle pathway in ⟨100⟩-oriented Ta nanocrystals subjected to uniaxial compression (a) and nanoindentation (b). (a1) Wedge-shaped twin comprised of {110} shuffle planes with staggered boundaries comprised of ⟨111⟩ symmetry lines or steps (a local {112} plane is highlighted in yellow). The prime directions *h*, *f* and *w* of twin growth are indicated. (a2) Cross-sectional view of an extended wedge-shaped twin that propagates from the surface to the center of the nanocrystal, along the *f*-direction. The ⟨100⟩ direction on which the critical zigzagged configuration emerges is highlighted at the HCP forefront of the platelet. The overall twin structure consists of twin variants "1" and "2" along with the region "3" where the HCP structure is about to instantly transform into variant "1". Moving boundaries between the HCP and BCC domains are highlighted with dashed lines. (a3) Schematic showing the capacity of the dual-shuffle process to form either straight {112} or staggered twin boundaries depending on the organization the ⟨111⟩ surface steps. (a4) Mixed platelet configuration arising in a ⟨140⟩-oriented nanocrystal, where the dual-shuffle and classic twinning modes operate concurrently on intercepting facets under the externally applied compression. (b) Dual-shuffle twinning dominates nanocontact plasticity. (b1) Early nucleation of lenticular twins underneath the indented surface where the staggered surfaces evolve into low-energy {112} twin boundaries. (b2) Cross-sectional view of the nanoindentation core region. c) Surface dislocation loop nucleation (dislocation line $\zeta$) in a tensile loaded ⟨111⟩ Ta nanocrystal facilitated through the dual-shuffle pathway.

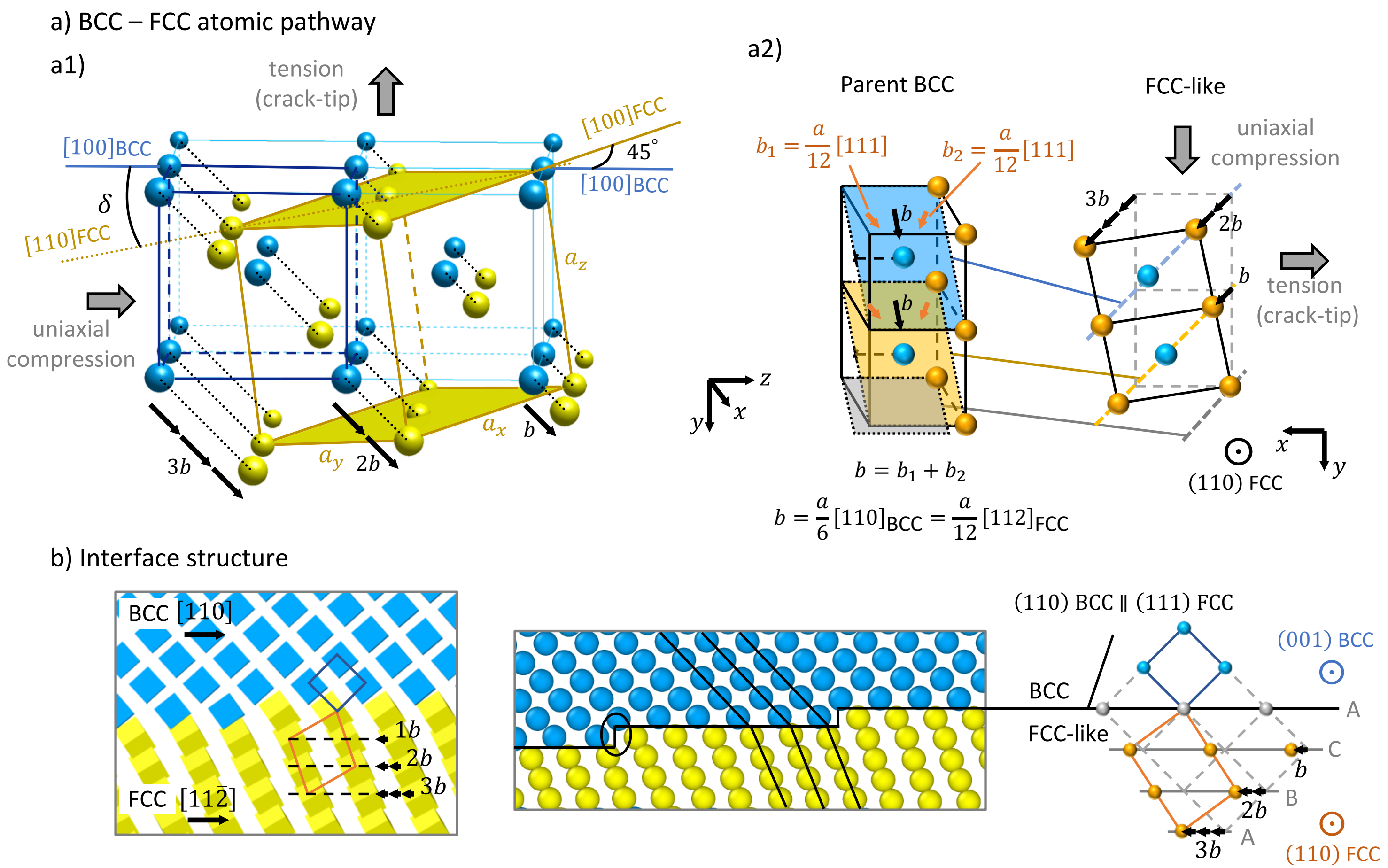

**Figure 4:** Atomic pathways for the formation of transient FCC-like structures (W tabGap potential). (a1) Nishiwama-Wasserman orientation relationship between the parent BCC (blue) and the derivative FCC-like (yellow) phases. The schematic representation comprises four BCC cells on which the $45^{\circ}$ axis rotation and $\delta = 10^{\circ}$ lattice rotation occur. The lattice rotation emerges from the incremental slip associated with collective partial dislocation glide processes each carrying a net Burgers vector $b$. (a2) Dislocation processes for the BCC-FCC transition. Partial dislocations glide on the marked $\{110\}$ planes of the parent BCC phase under the action of the applied shear stress. The arrows indicate the Burgers vectors $b_1$ and $b_2$ of the constituent $\langle 111 \rangle$ partial dislocations, producing the aforementioned net Burgers vector $b = b_1 + b_2$ along with incremental displacements on the $\{110\}$ planes. (a3) Coherent interfacial arrangement between FCC-like and parent BCC regions. Notice the distinctive lattice rotation and Burgers vectors of the partial dislocations (consistent with the sketches in panels (a1) and (a2)), along with the stepped character of the interphase that enables its fast mobility.

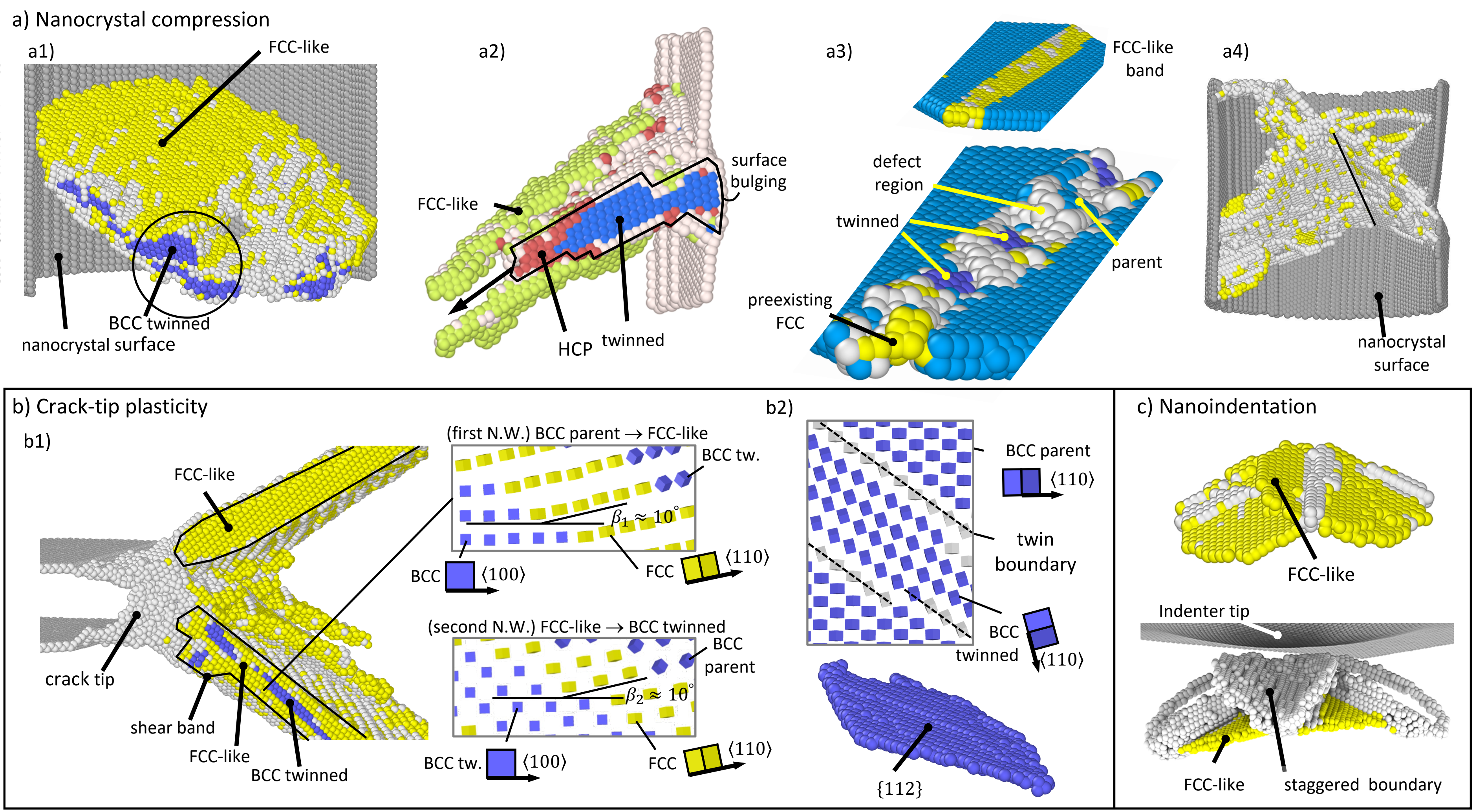

**Figure 5:** Emergence of twinned BCC structures from transient FCC phases. (a) Uniaxial compression of W nanocrystals (tabGAP potential). (a1) Surface nucleated platelets comprised of FCC-like boundary layers and inner twinned BCC regions produced under the dual-shuffle pathway. (a2) Detail of the surface defects leading to platelet formation in (a1). Notice the onset of the three layer structure mediated by the formation of the transitional HCP phase at the forefront which propagates in the direction of the arrow. (a3) Annihilation process of the FCC-like platelet in (a1) at the nanocrystal interior. Notice the onset of regions of high defect density (white) and twinned BCC sites (blue). (a4) Intercepting array of surface nucleated FCC-like platelets induced under tensile load application along ⟨111⟩ directions (platelet interception occurs along the black line). (b) Nucleation of FCC-like bands and twinned regions from crack tips (Mo-EAM and W-tabGAP potentials). The insets to Fig. (b1) indicate the characteristic angles $\beta_1$ and $\beta_2$ confirming the onset of two Nishiwama-Wasserman (N.W.) transformations on different {110} planes of the transitional FCC phase (see text for details). (b2) Final twinned structure distinguished by the formation of low-energy {112} boundaries. (c) Activation of the FCC-mediated twinning pathway during nanoindentation of (100) W surfaces (tabGAP potential). Notice the nucleation of FCC platelets followed by the emergence of twinned BCC regions through the dual shuffle process.

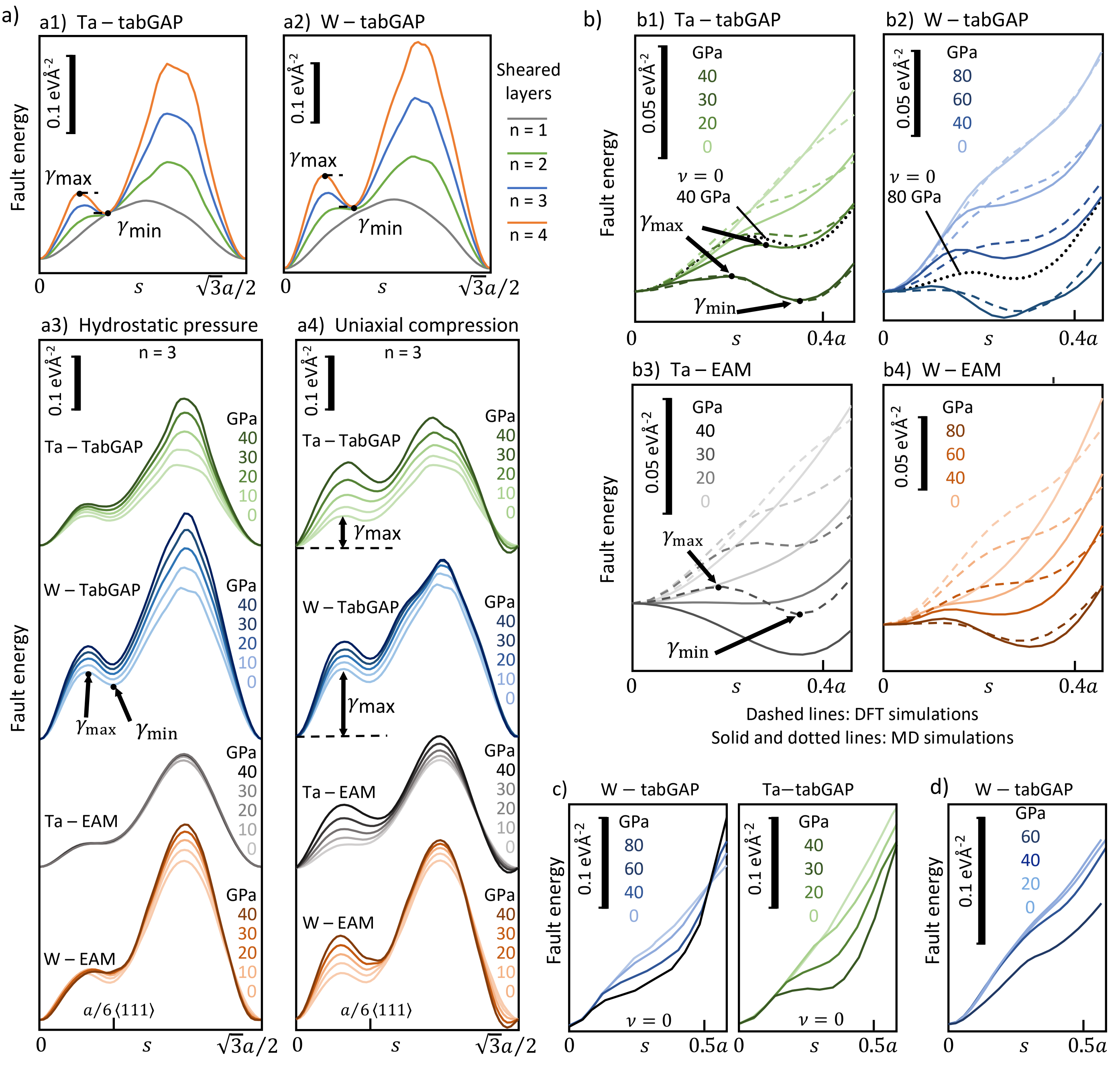

**Figure 6:** Energy landscapes for the twinning routes in archetypal Ta (low elastic stiffness) and W (high elastic stiffness) superlattices, modeled though EAM and Tab-Gap potentials, and associated DFT simulations. (a1) and (a2) MD simulations of the fault energy ($\gamma$) under the classic twinning pathway produced by the action of a shear displacement $s$ on a number of $n$ parallel $\{112\}$ planes displaced along $\langle 11\bar{1}\rangle$ directions. Notice the onset of the energy barrier $\gamma_{\max}$ and the relaxation energy $\gamma_{\min}$. (a3) and (a4) Uniaxial compression and hydrostatic pressure inhibit classic twinning through an increase in $\gamma_{\max}$. (b1) to (b4) Energy landscapes for the onset of the transitional HCP phase in the dual-shuffle pathway on infinite $\{110\}$ layers subjected to uniaxial compression. The shuffling distance $s$ is applied along $\langle 110\rangle$ directions (simulations under $\nu = 0$ indicate the influence of high hydrostatic components). Notice that the spontaneous onset of the HCP phase relies on the application of large compression, where both $\gamma_{\max} \to 0$ and $d\gamma/ds = 0$ at the vicinity of $s = 0$. (c) MD simulations of the energy landscapes for the formation of the HCP phase under high hydrostatic compression ($\nu = 0$) with *a single* $\{110\}$ shuffled layer. Notice that this process is essentially hindered due to the absence of energy minima. (d) MD simulations for the formation of FCC-like platelets on $\{110\}$ planes where the shear displacement $s$ is applied along <110> directions. Notice the inherent instability of the FCC-like phase which arises for $s = 0.25a$ under a gradual decrease of the curvature $d^2\gamma/ds^2$ in the absence of energy minima.

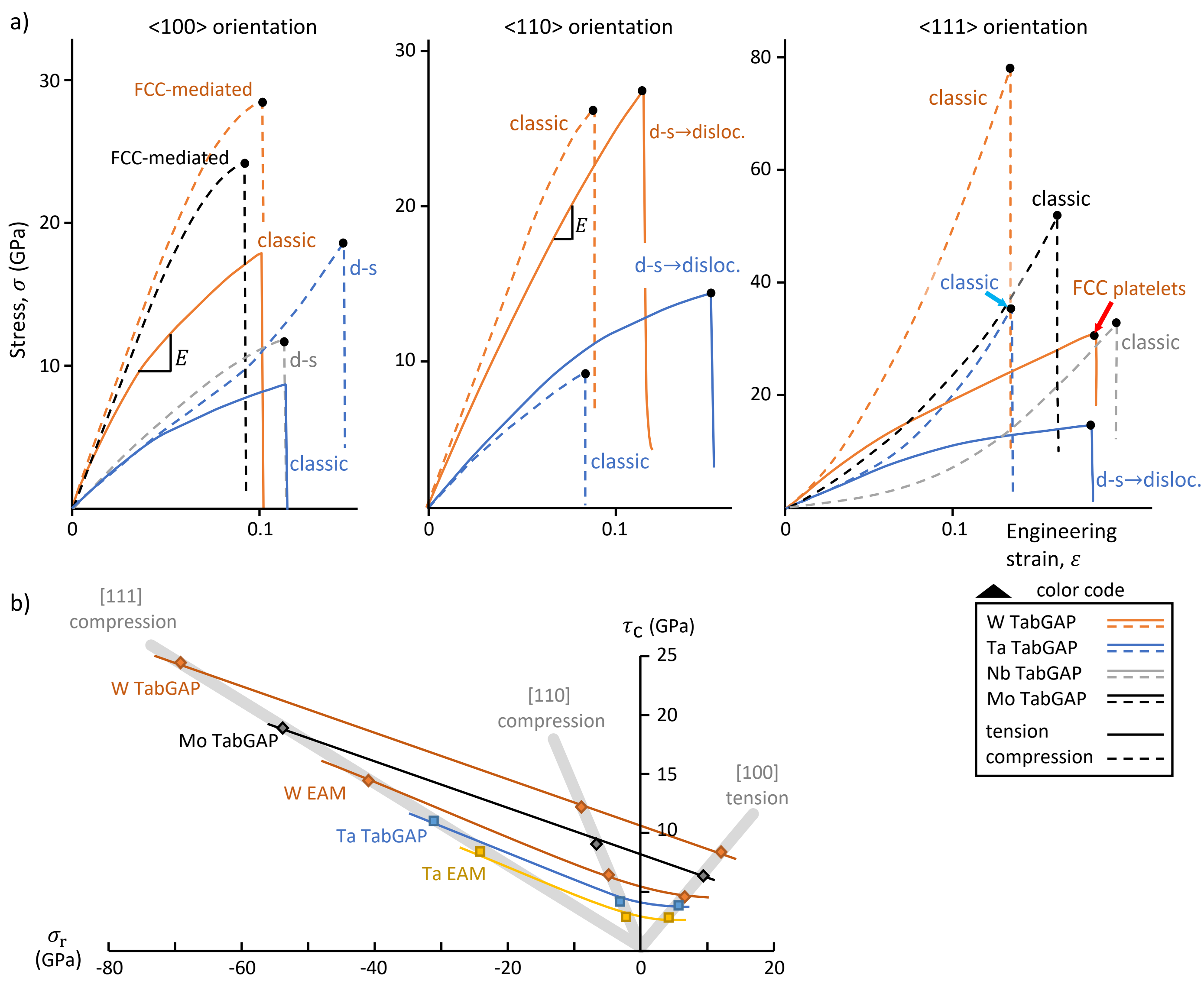

**Figure 7:** Stress-strain curves and pressure dependency of the critical shear stress for twinning in BCC nanocrystals. (a) Stress-strain curves for different loading directions where plastic instabilities occur at the critical pop-in stress $\sigma = \sigma_c$ marked with small black circles. The pop-in events indicated as "cl", "d-s", and "FCC-mediated" denote the activation of the classic, the HCP- mediated (dual-shuffle), and FCC-mediated twinning pathways. Pop-ins marked as "d-s→disloc." denote the onset of the dual-shuffle pathway from which dislocation nucleation is instantly triggered. The pop-in labelled "FCC platelets" indicates a plastic instability triggered by the formation of intercepting FCC-like platelets. (b) Critical shear stress $\tau_c$ for the activation of the *classical twinning route* (resolved on the critical $\{112\}\langle 11\bar{1}\rangle$ systems), as a function of the uniaxial compressive (−) or tensile (+) stress $\sigma_r$ acting on the $\{112\}$ twin boundaries. Thick grey lines indicate data sets from nanocrystals subjected to $\langle 111\rangle$ (compression), $\langle 110\rangle$ (compression) and $\langle 100\rangle$ (tension).

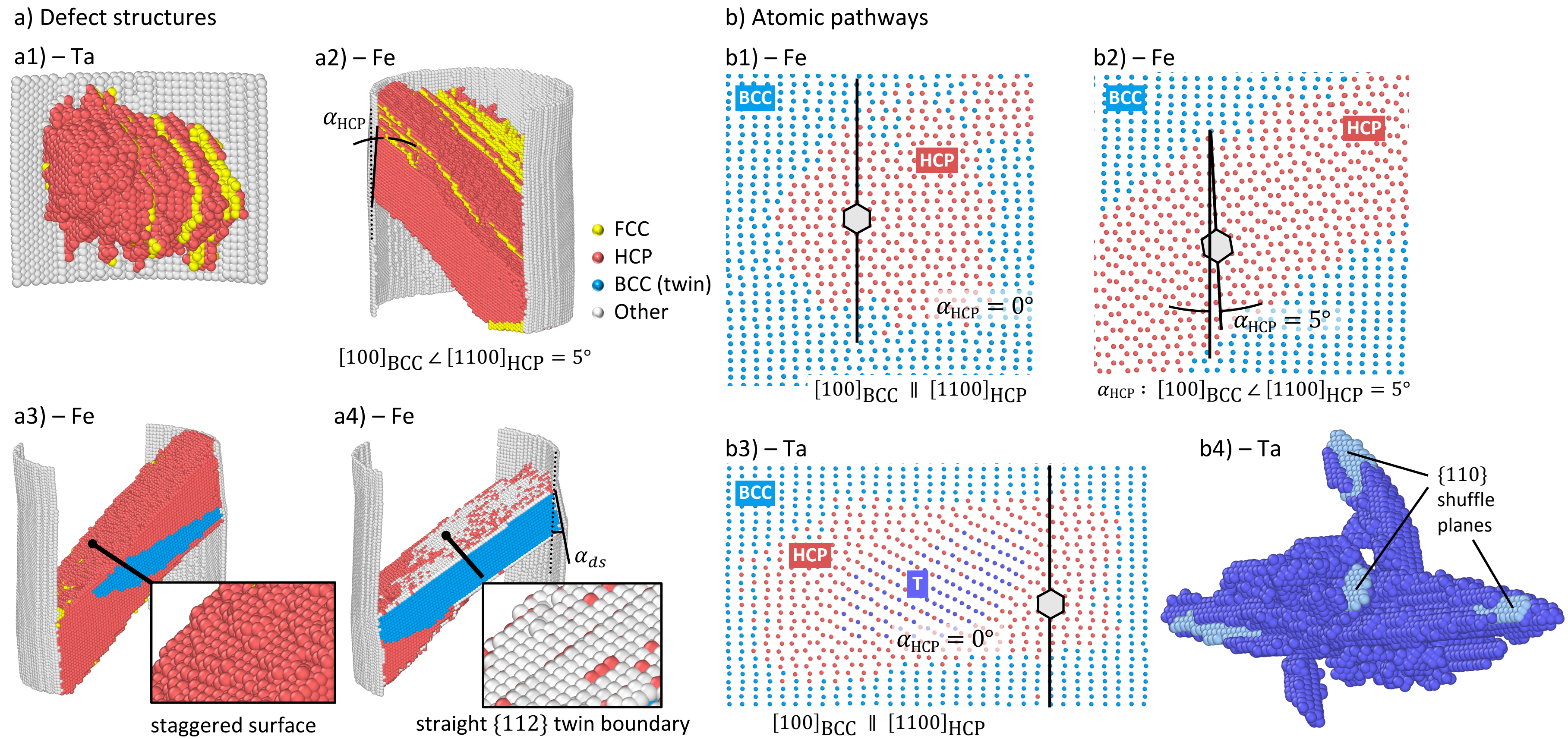

**Figure 8:** Influence of the thermodynamic stability of the transitional HCP phase upon the dual-shuffle twinning pathway (simulations under <100> compression). (a1) Twinning process in a free-standing Ta nanocrystal (EAM potential) mediated by the formation of a *metastable* wedge-shaped HCP phase that instantly transforms into the twinned BCC configuration at the pop-in stresses ($\sigma \rightarrow \sigma_c$). Panels (a2) to (a4) illustrate the progressive twinning process operating in an Fe nanocrystal, mediated by the formation of a *stable* HCP shear-band which transforms into the twinned BCC configuration as the stress is reduced during the pop-in event, $\sigma \ll \sigma_c$ (Fe-EAM potential). The angle $\alpha_{\mathrm{HCP}} \approx 5°$ prescribing shear band rotation and the final angle $\alpha_{\mathrm{ds}}$ are both indicated. The evolution from the staggered into the low-energy straight {112} twin boundary configurations is showcased in (a3) and (a4). (b) Dual-shuffle pathway in bulk Fe and Ta crystals in the absence of free surfaces (Fe-EAM and Ta-tabGAP potentials). Panels (b1) and (b2) show the formation of a *stable*, unrotated HCP-Fe nanodomain in a parent BCC region whose propagation ultimately leads to shear band formation with $\alpha_{\mathrm{HCP}} = 5°$. (b3) Development of an unrotated *metastable* HCP-Ta domain, instantly transforming into the twinned (T) configuration ($\alpha_{\mathrm{HCP}} = 0°$) at the pop-in stress ($\sigma \rightarrow \sigma_c$). (b4) Growth of the twinned nucleus in (b3) across the modelled supercell.

**Table 1:** Summary of transformation-mediated twinning processes

| | twinning mode | mechanisms | driving force |
|---|---|---|---|
| Lower elastic stiffness (Ta, Fe, Nb) | HCP-mediated<br>Uniaxial compression & nanoindentation preferentially along $\langle 100 \rangle$ directions | $BCC_p \rightarrow HCP \rightarrow BCC_t$<br>1st shuffle 2nd shuffle<br>1) Stable HCP phase (Fe): shear banding<br>2) Metastable HCP phase (Ta): wedge-shaped twins, platelets, and irregular nanodomains | $\langle 100 \rangle$ compression enables $\{110\}$ shuffles |
| Higher elastic stiffness (W, Mo) | FCC-mediated<br>Crack flanks | $BCC_p \rightarrow$ FCC—like $\rightarrow$ BCCt<br>double Nishiwama-Wassermann (N-W)<br>Nucleation of twinned BCC regions on FCC bands or platelets | resolved shear on $\{110\}$ planes & high tension |
| | FCC-mediated<br>Uniaxial compression near surfaces & nanoindentation | $BCC_p \rightarrow$ FCC platelets which triggers $BCC_p \rightarrow BCC_t$<br>N-W dual-shuffle<br>Formation of layered (FCC & $BCC_t$) structures | resolved shear on $\{110\}$ planes & high compression |
| | Uniaxial compression at the nanocrystal interior | FCC platelets $\rightarrow$ defect clusters, $BCC_p \rightarrow$ BCCt<br>dual-shuffle<br>Platelet annihilation produces twin pockets | |

Supplementary Material to **Transformation-mediated twinning governs plasticity in body-centered cubic nanocrystals under extreme loading**

Jan Očenášek, Jesper Byggmästar, Guanying Wei,
Javier Domínguez, and Jorge Alcalá

### *1. Crystallographic assessments of twin formation*

The classical twinning route involves the formation of $\{112\}$ twin boundary planes, which can be readily identified in MD simulations. A given $\{112\}$ twin boundary contains a unique *perpendicular* $\{110\}$ *P*-plane along with the ABAB… stacking sequence arises in the absence of any out-of-plane displacements. This perpendicular *P*-plane satisfices the necessary condition that the trajectory of any atom located in an A-layer (or B-layer) is consistently reproduced in the subsequent A-layer (or B-layer). In the dual-shuffle pathway, however, the formation of $\{112\}$ twin boundaries is generally precluded. The identification of the *P*-plane must therefore rely on a detailed tracking of atomic trajectories to ensure that the displacements across A and B-layers are consistently matched. This procedure is necessary because three of the six $\{110\}$ planes containing the characteristic $\langle 111\rangle$ symmetry line along which the net twinning shear is directed exhibit vanishing out-of-plane components, while only one of these three candidate *P*-planes satisfices the requirement of matching atomic displacements across A and B-layers. As illustrated in Fig. S1, the correspondence angle $\theta_{ds}$ can be uniquely measured in this *P*-plane, whereas it cannot be determined in any other $\{110\}$ plane fulfilling the condition of vanishing out-of-plane displacements.

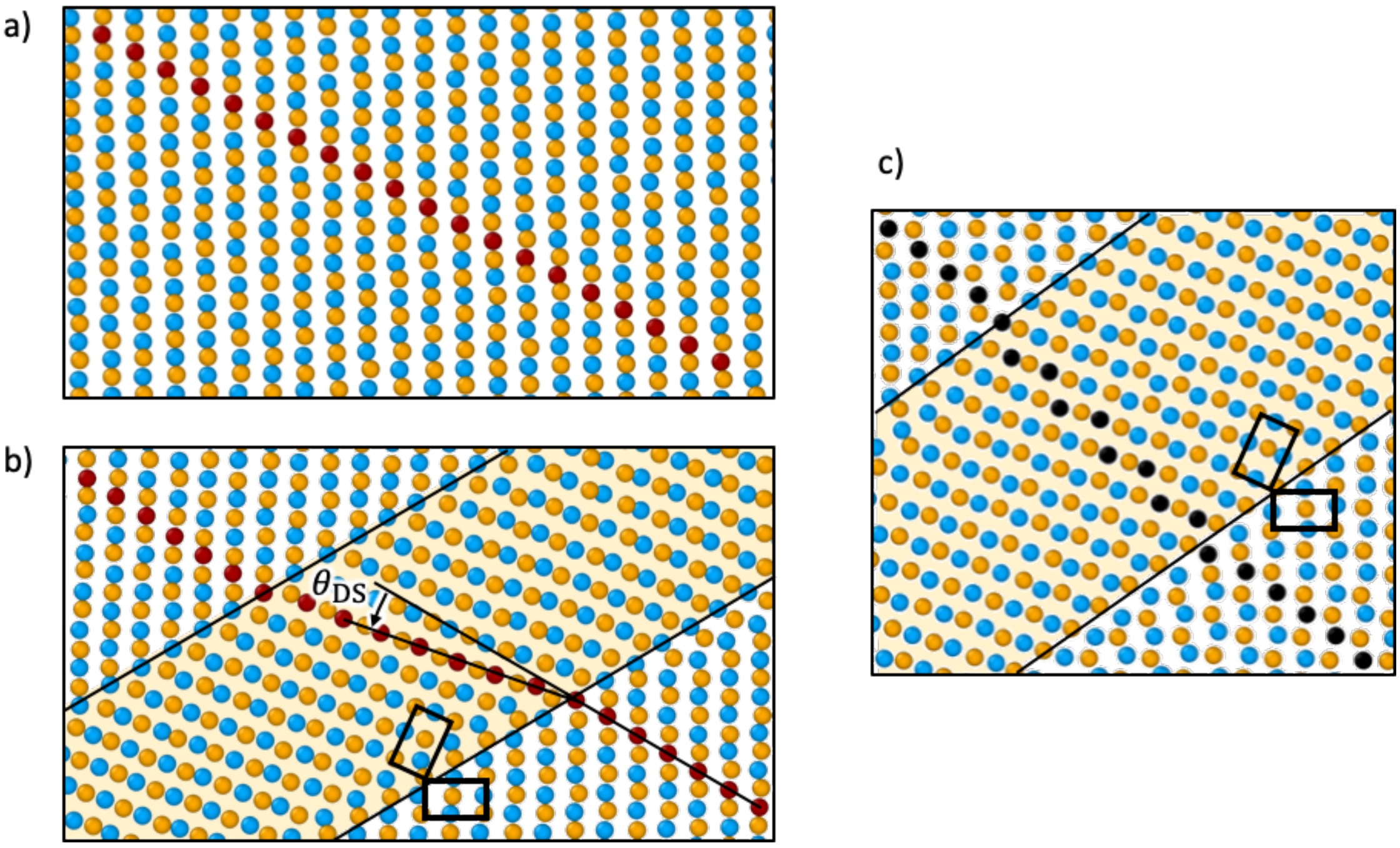

Fig. S1: Analysis of the correspondence angle $\theta_{\mathrm{ds}}$ for the dual-shuffle process (Ta-TabGAP potential). (a) Parent crystal in the original configuration prior to twin formation. Atoms in orange and blue indicate the A and B positions in the prevailing ABAB… stacking sequence of the parent crystal across the $\langle 110\rangle$ out of the paper direction. Notice that the straight line, marking the location of the darker atoms, points towards a $\langle 111\rangle$ direction. (b) Twinned region traversing

across the parent crystal along the same $\langle 110\rangle$ cross-sectional representation as in panel (a). The visualized, out of the paper, $(110)$ plane becomes the unique *P*-plane of the twinning process where the atomic trajectories across the ABAB… stacking are maintained, which ensures the correct measurement of the angle $\theta_{ds}$. (c) Analysis of the same twinned region as in panel (b) where the crystal is now visualized along a different $\langle 110\rangle$ cross-section (different to the *P*-plane in (b)) where the atomic trajectories across the ABAB… stacking sequence are not preserved. Notice the disruption of the atomic positions along the aforementioned $\langle 111\rangle$ direction, producing a zigzagged atomic arrangement that conceals the correct correspondence angle $\theta_{ds}$.

It shall be noted that the final atomic positions in the twinned lattice produced by the dual-shuffle process conceptually coincide with those rendered by the subsequent application of anti-twinning displacements with $b = 2a/6\ \langle\bar{1}\bar{1}1\rangle$ followed by twinning displacements with $b = a/6\ \langle 11\bar{1}\rangle$ on adjacent $\{112\}$ planes.

***2. Experimental assessment of wavy twins and phase transitions in compressed nanocrystals***

The observation of staggered twin boundaries in the dual-shuffle twinning pathway challenges conventional TEM analyses that conceal the three-dimensional twin boundary morphology. To this end, shear-driven twin formation processes in Ta polycrystals conducted with a Kolsky bar as well as in drawn samples illustrate the development of embryonic, multifaceted twin islands (Ref. [1]), and bulged twin boundaries (Ref. [2]). Although the twin structures observed in these experiments resemble the staggered twin morphologies arising in our simulations, the former develop through cooperative dislocation glide process on habit $\{112\}$ planes under small applied compression and hydrostatic pressure. The formation of wavy twin boundaries at *micrometer* scales through such shear-driven mechanisms depicts a clearly distinct scenario than that by which staggered twin boundaries arise under the dual-shuffle pathway in nanocrystals subjected to high compressive stresses.
High-pressure shock-loading experiments in Fe (Ref. [3]) and Nb (Ref. [4]), offer insights into twin formation processes mediated by transient (stable) hexagonal $\varepsilon$ and $\omega$ phases, respectively. While the potential of the transient HCP phase in Fe to exhibit crystalline rotations about its basal plane was considered in Ref. [3], the shear-driven nature of the assumed atomic processes leading to twin formation differs from the currently revealed dual-shuffle pathway triggered both under uniaxial compression and hydrostatic loadings (e.g., see Section 2.4 in the main text). Our simulations further show the central role of the rotated HCP phases upon shear band formation, providing an alternative perspective into the origin of the experimentally observed lenticular twin structures (Section 2.5 in the main text).
Along similar lines, it shall be noted that since the formation of the stable $\omega$ phase in the shock loading experiments of Nb in Ref. [4] involves much higher pressures than those applied in our uniaxial compression simulations, the experimentally observed twins consist of a frontal $\omega$ region mirroring the transient HCP phase in the dual-shuffle pathway. A clear distinction however emerges concerning the ultrafast propagation and thermodynamic instability of the transient HCP phase at the forefront of the platelets observed in our MD simulations, and the stability of the $\omega$ phase produced under shock loadings which enables its observation through post-mortem TEM analyses.
Following a shear-driven Nishiwama-Wasserman (N-W) orientation relationship (OR) similar to that identified in our MD simulations of the FCC-mediated twinning pathway (see Section 2.3 in the main text), a BCC-FCC-BCC phase transition was identified in the experiments in Ref. [5] concerning $\langle 100\rangle$-oriented Mo samples with crack flanks lying on $(100)$ planes. These results differ from other in situ

TEM observations illustrating the nucleation of stable FCC phases at regions of high stress concentration in $\langle 100 \rangle$-oriented Mo (Ref. [6]) and in $\langle 031 \rangle$-oriented Nb nanowires (Ref. [7]), which do not however transform back to the BCC phase. Notably, while derivative BCC phases indeed nucleated on the FCC regions ahead of crack tips (Ref. [5]), the ORs between the parent and derivative BCC configurations differed from those strictly enabling twin formation. The difference between the experiments and our MD simulations arguably arises from the abrupt unloading generated in the former upon nucleation of the transient FCC phase, whereas strict displacement-controlled loading conditions always prevailed in our MD simulations. Although these continuous loading processes ultimately triggered the nucleation of twinned BCC variants so as to lower the overall system energy, twin formation was circumvented in the experiments because of the decrease of the elastic energy induced during unloading. Hence, as the transitional FCC phase in the experiments became unstable in the unloading stage, a BCC variant with arbitrary OR naturally emerged in the absence of twin formation.

### *3. The onset of the ultrafast FCC-mediated twinning route in defect-free bulk BCC crystals*

Figure S2 illustrates the platelet structure and atomic pathways leading to the formation of twinned BCC regions via the FCC-mediated route discussed in Section 2.3 of the main text. These simulation results concern uniaxial compression loadings with BCC cells subjected to periodic boundary conditions, and show the progressive onset of an FCC band on $\{110\}$ planes. In the absence of free surfaces, this band nucleates at the center of the MD cell and spreads across the entire volume. As the system becomes progressively unloaded during the pop-in event, the FCC band annihilates and gives rise to regions of high defect density, as well as sites with the parent BCC configuration which ultimately transform into the twinned BCC structure via the dual-shuffle pathway. These mechanisms are fully consistent with those observed at the center of the compressed nanopillars in Section 2.3 of the main text.

### *4. Formation of FCC phases in defect-free bulk crystals during the dual-shuffle twinning pathway*

Our bulk MD simulations with supercells constructed under periodic boundary conditions illustrate the development of FCC and HCP bands or laminates, as illustrated in Fig. S3 for Fe. At the extreme applied compression triggering the initiation of plasticity in the absence of free surfaces, the formation of this *stable* FCC phase hinges on the similarity between the $\{0001\}$ planes in the HCP and the $\{111\}$ planes in the FCC lattices, enabling the first shuffle event from the parent BCC phase to produce either the ABAB… (HCP) or ABCABC… (FCC) packing sequences. The second shuffle event into the twinned BCC configuration is then subsequently attained upon the thermodynamic instability enabled upon substantial unloading ($\sigma < \sigma_c$), regardless of whether the transient phase has HCP or FCC packings (Fig. S3). The simulations further illustrate that the formation the FCC phase is facilitated at the vicinity of the HCP regions produced during the first shuffle event, ultimately leading to the formation of a thermodynamically stable composite structure containing crystallographic FCC variants under stress, which ultimately transforms into the twinned BCC structure upon substantial unloading. In contrast, the simulations performed with BCC Ta MD supercells using both tabGAP and EAM potentials show the prevalence of irregular *metastable* HCP clusters (rather than bands) produced during the first shuffle event (Fig. S4). Stacking fault formation is then observed under the extreme compression applied to these superlattices, resulting in an HCP to FCC phase transition. The metastable mixed HCP/FCC clusters produced through these processes ultimately evolve into the twinned BCC configuration during the second shuffle event at high stress ($\sigma \rightarrow \sigma_c$), as substantial unloading is unnecessary to facilitate phase instability.

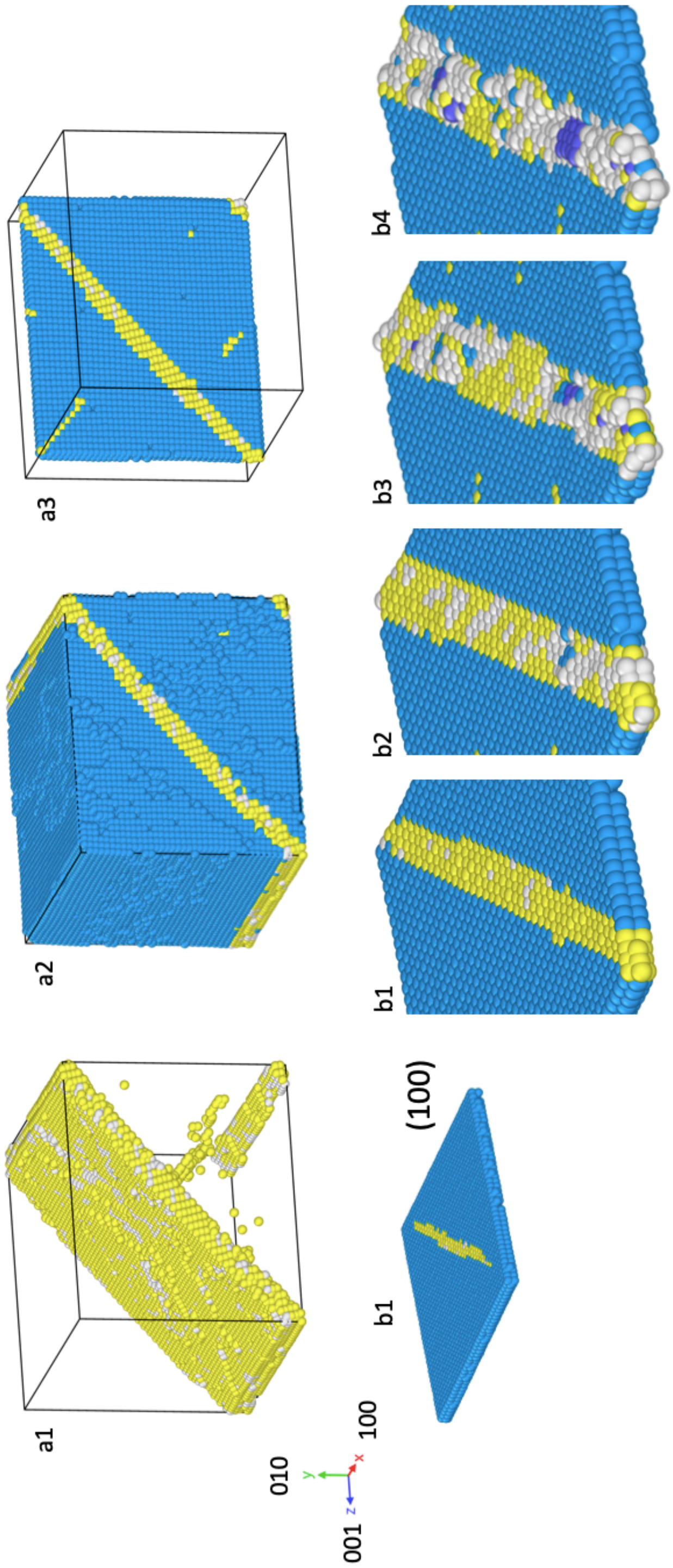

Fig. S2. Ultrafast FCC-mediated twinning pathway in the absence of free surfaces. The simulations are performed with computational cells subjected to periodic boundary conditions (W-tabGAP potential). Compression is applied along de $[010]$ direction, and leads to the nucleation of the FCC-like $(011)$ band shown in yellow in (a1), (a2), and (a3), where the parent BCC phase is highlighted in blue. The panel on the bottom shows a $[100]$ cross-sectional slice of the $(011)$ FCC band (b1). The snapshots in (b1) to (b4) showcase the progressive annihilation of the FCC phase, giving rise to regions with high defect density exhibiting clear out-of-plane displacements (in white). The parent BCC configuration selectively nucleates around these regions, ultimately transforming into twinned BCC sites (atoms marked in dark blue in (b4)).

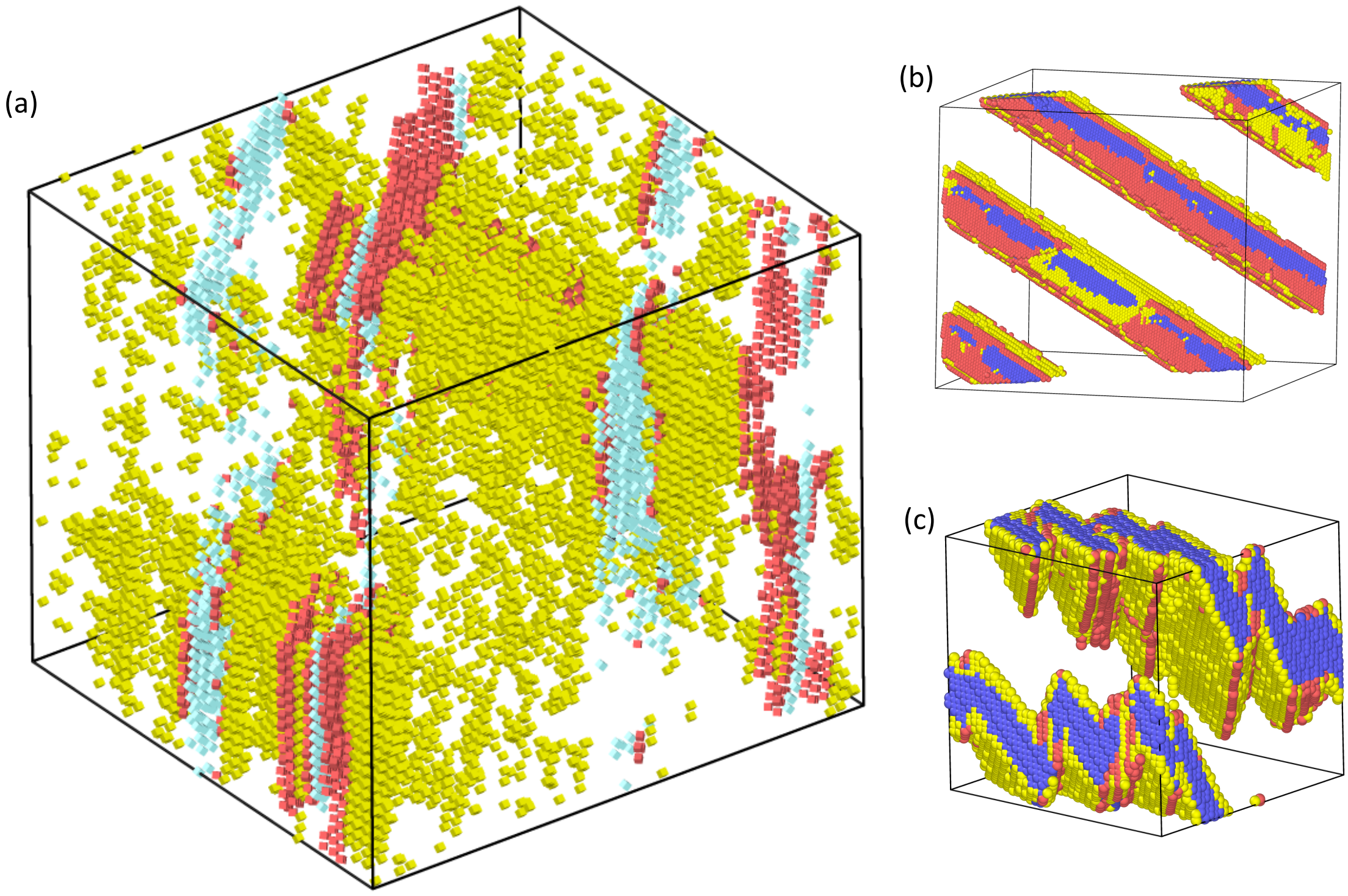

Fig. S3: Simulations of the dual-shuffle process with Fe supercells in the absence of free surfaces (EAM potential). (a) Nucleation of HCP and FCC regions on different locations of the modelled supercell, showcasing the orientation of the cuboidal unit cells. The formation of two FCC variants is highlighted in yellow and light blue, where the latter is preferentially located within the preexisting HCP laminates (in red). (b) Occurrence of the second shuffle event into the twinned configuration (in dark blue) on preexisting bands comprising both HCP (red) and FCC (yellow) regions. (c) Final lamellar structure produced as the stress is reduced during the plastic instability ($\sigma < \sigma_c$). The structure consists of truncated twinned BCC bands within conventional $\{112\}$ boundaries (in yellow).

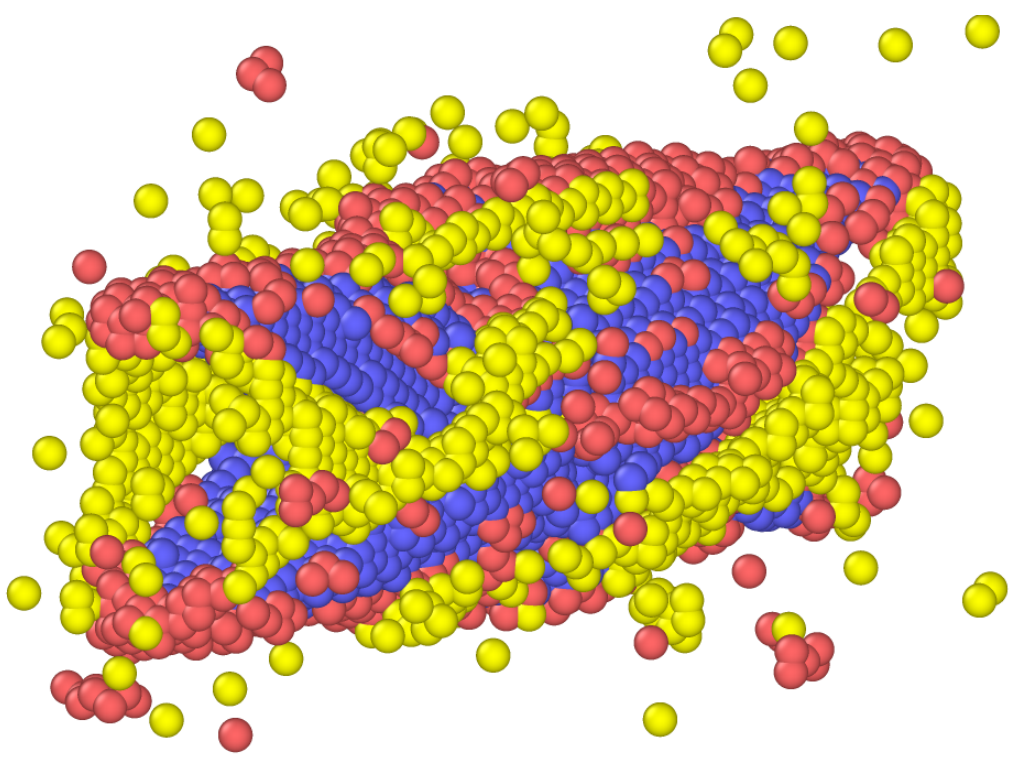

Fig. S4: Cross-sectional view of an irregular cluster produced during the dual-shuffle process within Ta supercells in the absence of free surfaces (tabGAP potential). Notice the alternated

arrangement of metastable HCP (red) and FCC (yellow) bands on which the second shuffle event into the twinned BCC structure (blue bands) progresses at $\sigma \to \sigma_c$.

### *5. Accuracy of the interatomic potentials in the MD simulations compared to DFT simulations*

The training dataset for the tabGAP potentials consists of a large number of atomic configurations arising in DFT simulations of BCC supercells of pure metals and alloys in various crystal structures and deformation states [8]. Each point in Fig. S25 represents the energy of a pure metal structure from the training set modeled supercell as a function of the calculated mean hydrostatic pressure ($p = (\sigma_{11} + \sigma_{22} + \sigma_{33})/3$), incorporating a rich variety of individual lattices with vanishing and non-vanishing normal stresses $\sigma_{11}$, $\sigma_{22}$ and $\sigma_{33}$. This is therefore indicative of the myriad of individual crystal lattices that were employed in the training of tabGAP potentials for the relevant interval of -25 GPa $< p <$ 25 GPa within which twin nucleation occurred in our MD simulations of nanocrystals subjected to simple tension and compression. Figure 6(b) in the main text illustrates the accuracy of the estimated fault energy when the BCC lattices are subjected to the specific loading trajectory that triggers the dual-shuffle twinning route. It shall be noted that the gradual reduction of the slope $\mathrm{d}\gamma/\mathrm{d}s$ at the vicinity of $s = 0$ in our simulations with tabGAP potentials reproduce the critical compression levels of ≈40 and ≈80 GPa for which the zigzagged atomic displacements into the HCP structure spontaneously occur in the DFT simulations of Ta and W, ultimately ensuring that the condition $\mathrm{d}\gamma/\mathrm{d}s < 0$ is enforced for increasing values of $s$. Substantially larger differences between the $\gamma - s$ curves from DFT and MD simulations however arise with EAM potentials (see Fig. 6(b)).
Figures S6 and S7 highlight the enthalpy evolution with increasing hydrostatic pressure obtained in the simulations with tabGAP and EAM potentials, as compared to those from DFT simulations. Clearly, the lower enthalpy reached for the BCC phase in the tabGAP and DFT simulations indicate the lack of any stable phase transition into FCC or HCP phases. The difference between the enthalpy-pressure curves for the BCC, FCC and HCP phases is however reduced in the simulations with EAM potentials, where unrealistic BCC to HCP and BCC to FCC transitions ultimately arise in W at hydrostatic pressures in the range of 30 GPa.
In the light of the ongoing discussion, it shall be noted that while the EAM potentials yield qualitatively accurate or phenomenologically correct results for the dual-shuffle route, the increased stability of the HCP phase that is particularly attained with the W-EAM potential leads to pronounced twin formation processes under the dual-shuffle pathway. This feature is clearly inconsistent with the advent of the FCC-mediated twining route predicted with the landmark tabGAP potentials for the elastically stiffer W and Mo nanocrystals.

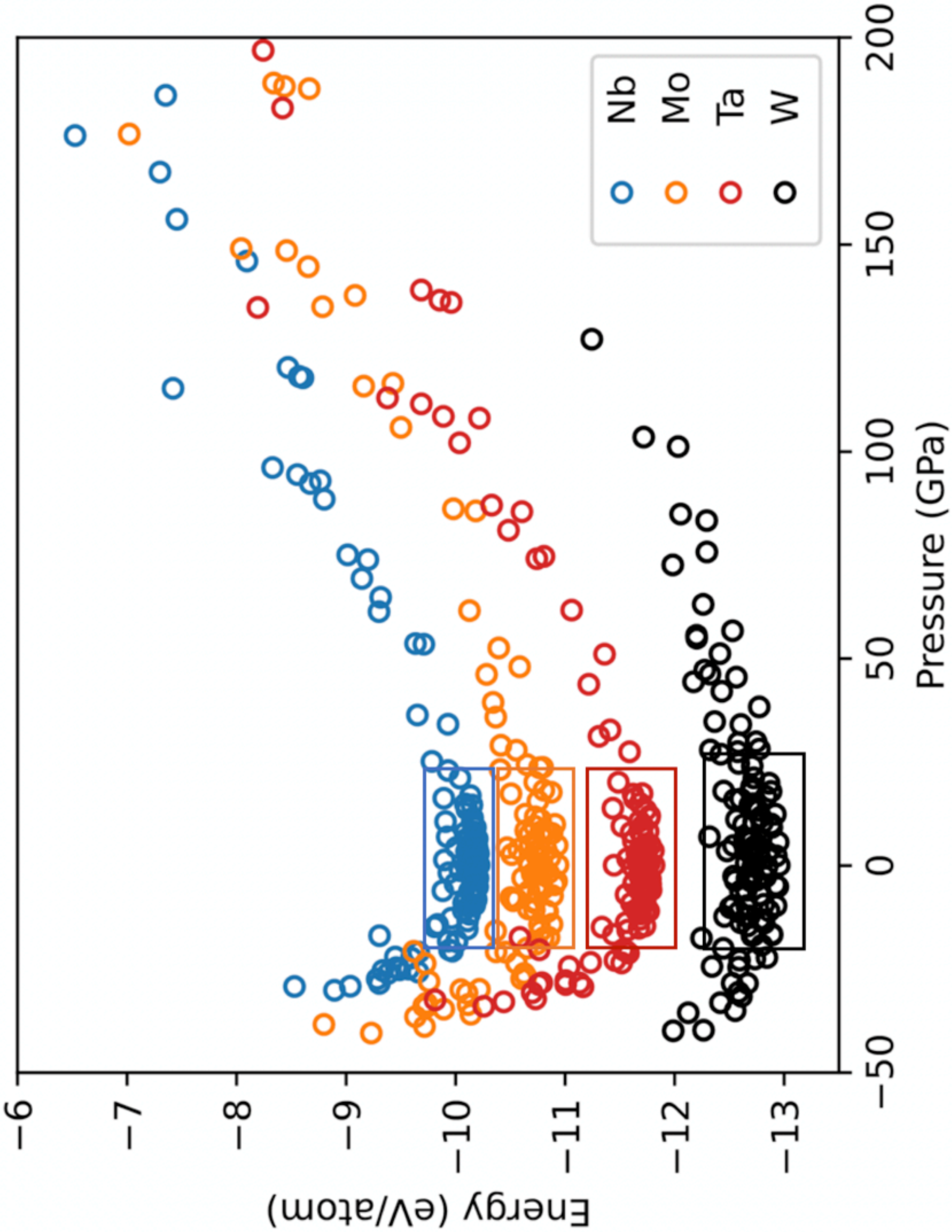

Figure S5: Energy of the training data structures for the tabGAP potentials as a function of hydrostatic pressure.

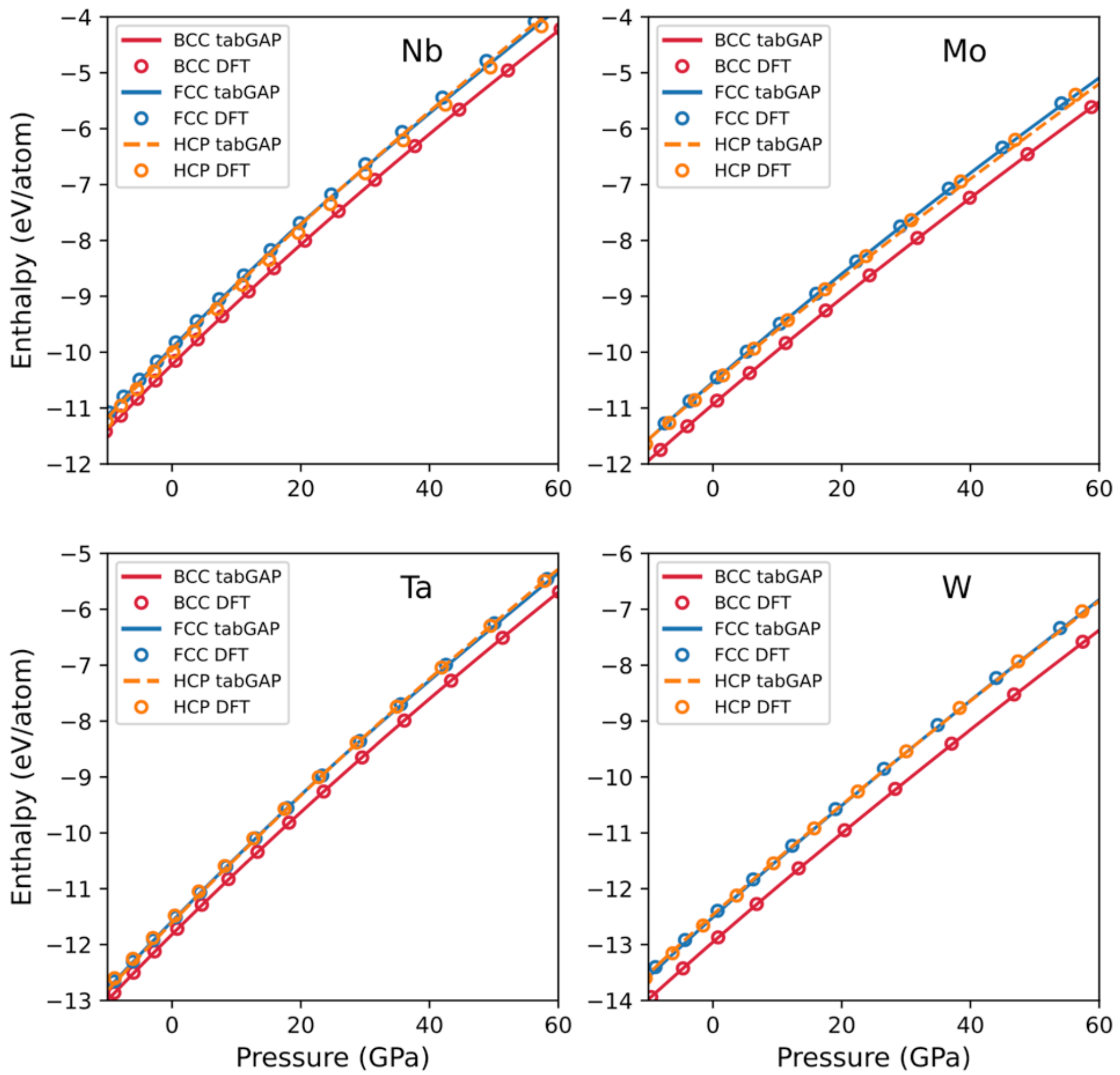

Figure S6: Enthalpy as a function of pressure for BCC, FCC, and HCP lattices in DFT simulations and MD simulations with tabGAP potentials.

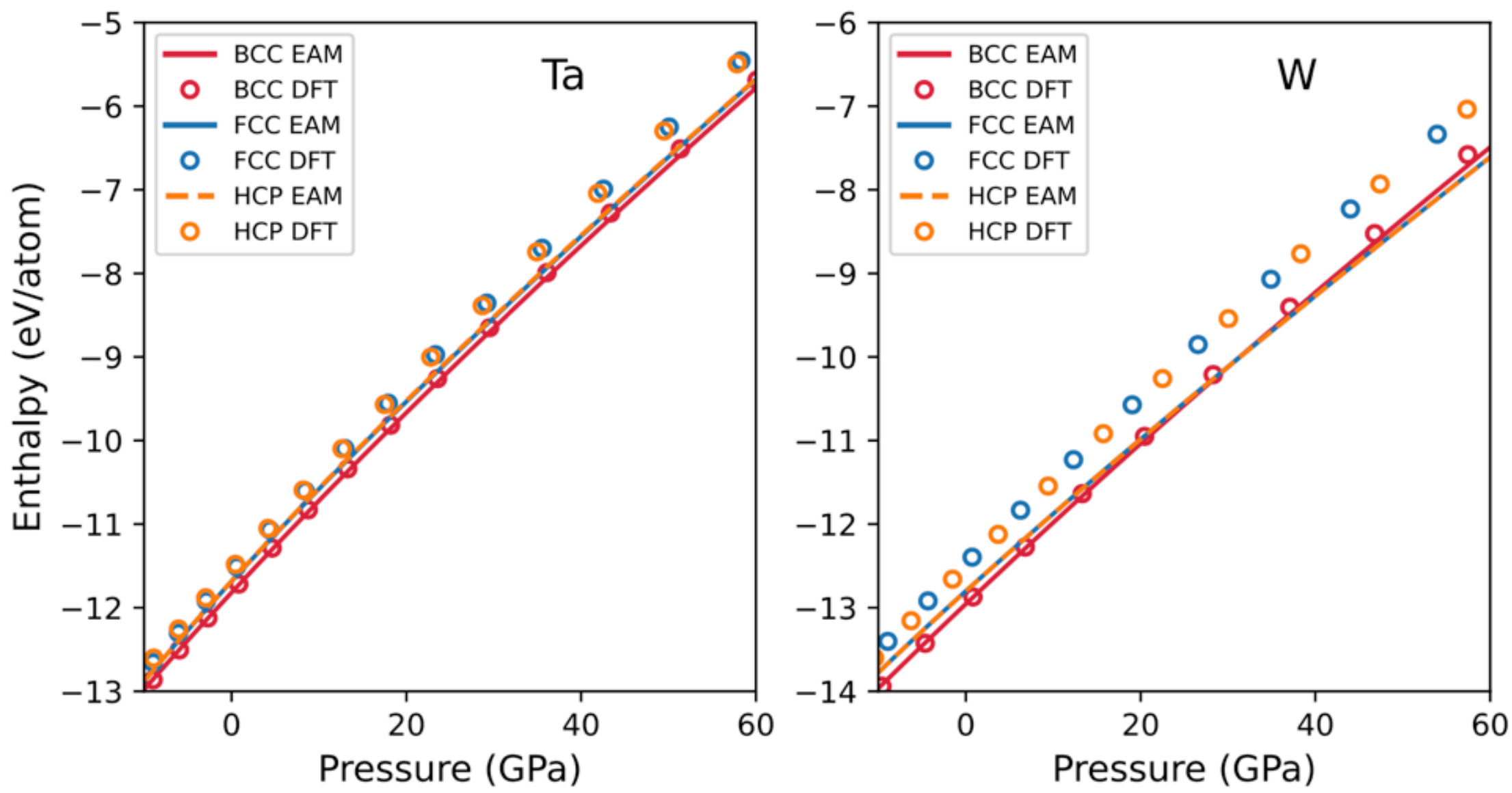

Figure S7: Enthalpy as a function of pressure for BCC, FCC, and HCP lattices in DFT simulations and MD simulations with EAM potentials (Ta, EAM; W, EAM)